%
\input phyzzx
\hfuzz 20pt
\font\mybb=msbm10 at 12pt

\def\Bbb#1{\hbox{\mybb#1}}

\def\bR{\Bbb {R}}
\def\bE{\Bbb {E}}

\def\bZ {\Bbb{Z}}
\def\bH {\Bbb{H}}
\def\bfomega{\omega\kern-7.0pt \omega}

\def\C{\mkern1mu\raise2.2pt\hbox{$\scriptscriptstyle|$}\mkern-7mu{\rm C}}

\def\cJ {{\cal {J}}}

\def\cL {{\cal {L}}}

\def\cMe {{\cal {M}}_{(8)}}



\REF\pktone{P.K. Townsend, {\sl The eleven-dimensional 
supermembrane revised}, Phys. Lett. {\bf 350}
(1995) 184 [hep-th/9501068]. }
\REF\malda{J. Maldacena,{\sl The large N limit of superconformal field theories
and supergravity} [hep-th/9711200].}
\REF\gppkt{G. Papadopoulos and P.K. Townsend,  
{\sl Intersecting M-branes}, Phys. Lett.
 {\bf B380} (1996) 273 [hep-th/9603087].}
\REF\tseytlinone { A.A. Tseytlin, {\sl Harmonic 
superpositions of M-branes}, Nucl. Phys. 
{\bf B475} (1996) 149 [hep-th/9604035]}
\REF\jerome{J. Gauntlett, D. Kastor and
 J. Traschen, {\sl Overlapping branes in M-theory}, 
Nucl. Phys. {\bf B478} (1996) 544
[hep-th/9604179]. }
\REF\eric{ E. Bergshoeff, M. de Roo, E. Eyras, 
B. Janssen, J.P. van der Schaar, {\sl
Intersections involving monopoles and waves in eleven dimensions}, 
Class. Quantum Grav. {\bf 14} (1997)
2757 [hep-th/9704120]. }
\REF\bre{J.C. Breckenridge, D.A. Lowe, R.C. Myers, 
A.W. Peet, A. Strominger and C. Vafa, {\sl
Macroscopic and microscopic entropy of near-extremal 
spinning black holes} [hep-th/9603078].}
\REF\cvetic{M. Cveti\v c and F. Larsen, {\sl Near-horizon 
geometry of rotating black holes in
five dimensions} [hep-th/9805097]; {\sl Microstates of 
four-dimensional rotating black holes from
near horizon geometry} [hep-th/9805146].}
\REF\cvey{M. Cveti\v c and D. Youm, {\sl Rotating Intersecting M-Branes}, Nucl. Phys. 
{\bf B499} (1997) 253 [hep-th/9612229].}
\REF\jerometwo{J.P. Gauntlett, R.C. Myers and 
P.K. Townsend, {\sl Supersymmetry of rotating
branes} [hep-th/9809065].}
\REF\horoark{ G. T. Horowitz and A.A. Tseytlin, {\sl A new class of exact
solutions in string theory}, Phys. Rev. {\bf D51} (1995) 
2896 [hep-th/9409021].}
\REF\gtpap{J.P. Gauntlett, G.W. Gibbons, 
G. Papadopoulos and P.K. Townsend, {\sl Hyper-K\"ahler
Manifolds and Multi-Intersecting Five Branes},  Nucl. Phys.
{\bf B500} (1997) 133 [hep-th/9702202]}
\REF\pandrew{ G. Papadopoulos and A. Teschendorff, 
{\sl Multi-angle five brane intersections}, Phys. Lett. {\bf B443} 
(1998) 159 [hep-th/9806191]
; {\sl Grassmannians, Calibrations and 
Five-Brane Intersections} hep-th/9811034.}
\REF\ital{L. Castellani, A. Ceresole, R. D' Auria, 
S. Ferrara, P. Fre', and M. Trigiante,
{\sl G/H M-branes and $adS_{p+2}$ geometries} [hep-th/9803039].}
\REF\hull{B. S. Acharya, J.M. Figueroa-O' Farrill, 
C.M. Hull, and B. Spence,{\sl Branes 
at Conical Singularities and Holography} [hep-th/9808014].}
\REF\cowdall{P. Cowdall and P.K. Townsend, {\sl Gauge 
Supergravity Vacua from Intersecting branes }, 
Phys. Lett. {\bf B429} (1998) 281; Erratum-ibid {\bf B434} 
(1998) 458 [hep-th/9801165].}
\REF\boonstra{H. J. Boonstra, B. Peeters and K. Skenderis, 
{\sl Brane Intersections, Anti-de Sitter spacetimes and dual
superconformal theories}, hep-th/9803231.}
\REF\julia{B. Julia and H. Nicolai, {\sl Null-Killing 
Vector Dimensional Reduction
and Galilean Geometrodynamics}, Nucl. Phys. {\bf B439} 
(1995) 291 [hep-th/9412002].}
\REF\corr{ E. Corrigan, C. Devchand, D.B. Fairlie, 
J. Nuyts, {\sl First Order
equations for gauge fields in spaces of dimension 
greater than four}, Nucl. Phys. {\bf B214}
(1983) 452.}
\REF\ward{ R.S. Ward, {\sl Completely solvable gauge-field 
equations in dimensions greater
than four}, Nucl. Phys. {\bf B236} (1984) 381.}
\REF\kent{E. Corrigan, P. Goddard and A. Kent, 
{\sl Some Comments on the ADHM Construction
in 4k Dimensions}, Commun. Math. Phys. {\bf 100} (1985) 1.}
\REF\donaldson {S.K. Donaldson, {\sl Anti Self-dual 
Yang-Mills Connections over Complex
Algebraic Surfaces and Stable Vector Bundles}, Proc. 
London Math. Soc. {\bf 50} (1985) 1.}
\REF\popov{A.D. Popov, {\sl Anti-self-dual solutions of the 
Yang-Mills equations in $4n$ dimensions},
Mod. Phys. Lett. Vol {\bf 7} (1992), 2077.}
\REF\gibpap{G.W. Gibbons and G. Papadopoulos, {Calibrations 
and Intersecting Branes} [hep-th/9803163].}
\REF\gpandrew{G. Papadopoulos and A. Teschendorff, 
{\sl Instantons at Angles}, Phys. Lett. {\bf B419} 
(1998) 115 [hep-th/9708116].}
\REF\fn{ S. Fubini and H. Nicolai, {\sl The 
Octonionic Instanton}, Phys. Lett. {\bf 155B}
(1985) 369.}
\REF\salamon{S. Salamon, {\sl Riemannian Geometry 
and Holonomy Groups }, Pitman Research Notes
in Mathematical Series 201, UK.}
\REF\poon{H. Peterson and Y-S. Poon, Commun. 
Math. Phys. {\bf 117} (1998) 569.}
\REF\gibbons{G.W. Gibbons, {\sl The Sen conjecture for fundamental monopoles
of distinct types}, Phys. Lett. {\bf B382} (1996) 53 [hep-th/9603176].}
\REF\trisasak{C.P. Boyer, K. Galicki, B.M. Mann and E. Rees, {\sl Compact
3-Sasakian 7-manifolds with arbitrary second betti number}, Invent.
Math. {\bf 131} (1998) 321; R. Bielawski, {\sl Betti 
numbers of 3-Sasakian quotients
of spheres by tori}, MPI preprint (1997)}
\REF\saseins{L. Castellani, L.J. Romans and N.P. Warner, 
{\sl A classification of
compactifying solutions for d=11 supergravity}, 
Nucl. Phys. {\bf B241} (1984) 429.}
\REF\arkadytwo{N. Itzhaki, A. A. Tseytlin and 
S. Yankielowicz, Phys. Lett. {\bf B432}
(1998) 298 [hep-th/9803103].}
\REF\green{M.B. Green, J. Schwartz and E. Witten, {\sl 
Superstring Theory}, Vol II, Cambridge
University Press, London (1987)\ .}
\REF\paptown{G. Papadopoulos and P.K. Townsend, 
{\sl Compactifications of D=11 Supergravity
on spaces of exceptional holonomy}, Phys. Lett. 
{\bf B357} (1995) 300 [hep-th/9506150]. }
\REF\townsenda{P.K. Townsend, {\sl M-theory from its superalgebra},
 Cargese lectures (1997) [hep-th/9712004].}
\REF\hullb {C.M. Hull, {\sl Gravitational Duality, Branes and Charges},
Nucl. Phys. {\bf B509} (1998) 216 [hep-th/9705162].}
\REF\mukhi{K. Dasgupta and S. Mukhi, {\sl Brane 
Constructions, Conifolds and M-theory}
 [hep-th/9811139].}

\Pubnum{ \vbox{ \hbox{DAMTP-1999-30}\hbox{} } }
\pubtype{}
\date{February, 1999}
\titlepage
\title{Rotating Rotated Branes}
\author{ G. Papadopoulos}
\address{DAMTP,\break Silver Street, \break  
University of Cambridge,\break Cambridge CB3
9EW}
\abstract {We present 
 a class of spacetime
rotations that preserve a proportion of 
spacetime  supersymmetry. We then
give the  rules for superposing these  rotations 
with various branes to  construct rotating brane 
solutions which preserve  exotic fractions of supersymmetry. We also
investigate the superposition of rotations 
with intersecting branes at angles and we find  new rotating  intersecting
branes at angles configurations. We demonstrate this with 
two examples of such solutions 
one involving intersecting NS-5-branes
 on a string at $Sp(2)$ angles superposed with fundamental 
strings and pp-waves, and the 
other involving intersecting M-5-branes
on a string at $Sp(2)$ angles superposed 
with membranes and pp-waves.  We 
find that the
geometry  of some of these solutions near the intersection region of 
every pair of 5-branes is 
$AdS_3\times S^3\times S^3\times \bE$ and
$AdS_3\times S^3\times S^3\times
\bE^2$, respectively.  We also present a class of  
solutions that can be used for null
string and M-theory compactifications
 preserving supersymmetry.  }

\endpage
\pagenumber=2



\chapter{ Introduction}

Many of the insights in the non-perturbative structure of the 
superstrings and in the  understanding of M-theory
have been found by investigated the 
soliton-like solutions of the ten- and
eleven-dimensional supergravity theories [\pktone, \malda].  Most of the
attention so far has been focused on 
supersymmetric solutions, {\sl i.e.} 
those that they preserve
a proportion of the supersymmetry of the
 underlying theory. It is
remarkable that a large class of 
supersymmetric solutions of the supergravity theories 
can be constructed by superposing
\lq elementary' soliton solutions which preserve 1/2 of the
 spacetime supersymmetry [\gppkt, \tseytlinone, \jerome].  These elementary
solutions are the various brane solutions of
supergravity theories, the pp-wave and KK-monopole.
After such a superposition, the resulting solutions have the 
interpretation of intersecting branes or 
branes ending on other branes,
and whenever appropriate, in the
 background of a pp-wave or a KK-monopole [\eric].
Recently it has been realized that angular momentum can be
superposed to  brane or to intersecting brane configurations
in such way that a proportion of 
spacetime supersymmetry is preserved [\bre, \cvetic, \cvey, \jerometwo].
The resulting configurations have the 
interpretation of  rotating branes or  rotating intersecting
branes. Such backgrounds have also been consider in the context of sigma models
as exact solutions of string theory [\horoark].  

In this paper, we shall present a systematic 
investigation of the type
of rotations that can be added  to branes and 
intersecting branes to preserve
 a fraction 
of spacetime supersymmetry.
We shall find that in many cases supersymmetric 
rotations are associated with
solutions of Maxwell field and Killing spinor equations on a possibly 
curved background. In 
turn, supersymmetric solutions
of Maxwell field equations can be found by integrating certain 
BPS-like conditions in various dimensions which are 
associated to certain subalgebras of 
$so(8)$. A list of such subalgebras
includes $spin(7)$, $g_2$, $sp(2)$ and $su(n)$ for $n=2,3,4$.
We shall give the explicit solutions of the Maxwell
equations  associated with all these 
subalgebras. Then we shall consider
the associated spacetime solution.
We shall find that such  spacetimes are 
asymptotically flat with zero mass and
momentum but they have non-vanishing angular 
momentum.  Some of these
spacetimes could be thought as \lq elementary', for example 
those associated with the subalgebras
$spin(7)$ and $su(n)$ for $n=2,3,4$, and so 
they can be superposed with branes to give new solutions with the
interpretation of rotating branes.  
However unlike the \lq elementary' branes, pp-waves and KK-monopoles,
spacetimes with only \lq elementary' rotation may preserve less 
than $1/2$ of spacetime supersymmetry\foot{This  indicates that
there may be another interpretation for this rotations probably
in terms of more \lq elementary' objects. However we shall not
pursue this further here.}.
Some other rotations can be constructed by a non-parallel 
superposition of \lq elementary' ones.
Such rotating spacetimes are associated with $sp(2)$ and $g_2$ subgroups.
We shall find that these composite rotations can be superposed
to intersecting branes at angles. Thus we construct
solutions with the interpretation of rotating rotated intersecting branes.
In what follows, we
shall refer to intersections or superpositions
of branes
at $G$-angles  as $G$-intersections or $G$-superpositions,
respectively. We shall present two main examples of such 
rotating rotated intersecting
branes. These examples involve the 
$Sp(2)$-intersecting NS-5-branes and the
$Sp(2)$-intersecting M-5-branes on a string  of [\gtpap, \pandrew], 
respectively.
In addition, pp-waves will be added to both configurations as well
as strings in the former and  membranes in the latter.
Rotations will also be investigated
in the background of $Sp(2)$-superpositions of KK-monopoles  which
is described by a toric eight-dimensional hyper-K\"ahler metric. The 
rotation in this
case is related to the two tri-holomorphic vector
 fields of this geometry.
Rotations will be examined in the background of Ricci-flat
cones with special holonomy and their various 
superpositions with strings,
membranes and pp-waves will be also considered. In particular, this 
will add rotation to  the
solutions found in [\ital, \hull].
The proportion of spacetime supersymmetry 
preserved by all the above
configurations will also be given. We 
shall find that in many cases
the solutions preserve various exotic 
fractions of supersymmetry.

The geometries near the intersection region 
(near horizon) of the above rotating  
$Sp(2)$-intersecting 5-brane configurations 
 will be investigated. We shall  find that the geometry
near the intersection region of every pair of NS-5-branes involved in a
$Sp(2)$-intersection on a string 
is $\bE^{(1,3)}\times S^3\times S^3$. This geometry
is the same as that for the associated orthogonal 
intersection [\cowdall, \boonstra].
Then we shall find that the geometry at the intersection region
of rotating $Sp(2)$-intersecting NS-5-branes  superposed
with pp-waves and strings is $AdS_3\times S^3\times S^3\times \bE$.
In addition, the geometry at the intersection region
of rotating $Sp(2)$-intersecting M-5-branes  superposed
with pp-waves and membranes is $AdS_3\times S^3\times S^3\times \bE^2$.
For this in both cases, we shall present  approximate solutions 
which has the desirable behaviour near the intersection
regions. We remark that these near horizon geometries are the same
as those found in [\cowdall,\boonstra, \jerometwo] for the 
associated orthogonal intersections.
Finally we shall show that  spaces with 
topology $\bR^n\times {\cal L}$ solve
the field equations of D=11 and D=10 supergravities,
 where ${\cal L}$ is a principal
$U(1)$ bundle over either tori, or Calabi-Yau or 
other manifolds with special holonomy.
We shall argue that such spaces can be used to 
compactify strings and M-theory
along directions for which one is null. This 
adapts the null compactifications investigated for example in [\julia]
 to strings and M-theory.

This paper has been organized as follows:
In section two, we  present the ansatz for our solutions
in the context of type II strings.  In section three,
we investigate spacetimes with 
pure rotation and explain the
relation between rotations and 
subalgebras of $so(8)$. We then
present new solutions with the 
interpretation of rotating strings and rotating 
NS-5-branes. Superpositions of  
 pp-waves are also considered.
In section four, we investigate 
rotations in the background 
of toric hyper-K\"ahler
manifolds.  In section five, we 
construct new solutions with the
interpretation of rotating rotated NS-5-branes 
and then superpose them
with strings and waves. In section 
six, we investigate rotations
in the background of Ricci-flat cones with 
special holonomy and then
superpose them with strings and 
membranes.  In section seven, we give an M-theory
interpretation to our solutions by lifting them 
to eleven dimensions. In section eight,
we find the geometry of the rotating 
rotated 5-brane solutions near
the intersection of every pair of 5-branes involved in the 
configuration. In section nine, we find the 
compact solutions of strings and M-theory
that can be used for null compactifications. 
In section ten, we give
our conclusions and in Appendix A we investigate the 
type II T-duality and IIB S-duality
properties of our solutions.

\chapter{Common Sector Rotations }

The common sector fields or  the NS$\otimes$NS 
fields of type II  and heterotic string theories include
the spacetime metric $g$, a three-form field strength $H$
and the dilaton 
$\phi$. The 
field equations of the associated
effective supergravity theories can be 
consistently truncated to this sector and they are as follows:
$$
\eqalign{
R_{MN}-H_{MPQ} H_N{}^{PQ}+2 \nabla_M\partial_N\phi&=0
\cr
\nabla_P (e^{-2\phi} H^{PMN}\big)=0}\ ,
\eqn\feq
$$
where we have set all fermions to zero and $R_{MN}$ is the
Ricci tensor of the metric $g_{MN}$; $M,N,P,Q=0,\dots, 9$.
There is another field equation that of the dilaton but
it is implied by the above two equations (up to a constant).

The Killing spinor equations for type II theories that
 involve only the
above fields are
$$
\eqalign{
\nabla^{(\pm)}_M \epsilon_{\pm}&=0
\cr
\big(\Gamma^M\partial_M\phi\mp {1\over 3!} H_{MNP} 
\Gamma^{MNP}\big)\epsilon_{\pm}&=0\ ,}
\eqn\inone
$$
where $\epsilon_{\pm}$ are 16-component
 Killing spinors and the connections 
of the covariant derivatives
$\nabla^{(\pm)}$ are
$$
\Gamma^{(\pm)}{}^M{}_{NP}=\Gamma^M{}_{NP}\pm H^M{}_{NP}\ ;
\eqn\intwo
$$
$\Gamma^M{}_{NP}$ is the Levi-Civita connection. In IIA 
and IIB supergravities,
$\epsilon_+$ and $\epsilon_-$ have either 
the opposite or the same chirality, respectively.
For the heterotic string, we simply 
truncate the above four Killing spinor equations
to two which are those of either	 $\epsilon_+$ or $\epsilon_-$.

We shall seek solutions that describe the superposition
of rotations with configurations of the common sector.
These configurations include the fundamental 
string, the NS-5-brane,
the pp-wave and the ten-dimensional KK-monopole as well as
their intersections or superpositions.
 To describe the ansatz of such configuration, we begin with a static 
solution on ${\cal M}_{(10)}=\bE^{(1,1)}\times {\cal M}_{(8)}$
  of the NS$\otimes$NS sector
 of the form
$$
\eqalign{
ds^2&=ds^2(\bE^{(1,1)})+ds^2_{(8)}
\cr
H&=H_{(8)}
\cr
e^{2\phi}&=e^{2\phi_{(8)}}\ ,}
\eqn\inthree
$$
where 
 $$
\eqalign{
ds^2_{(8)}&= \gamma_{ab} dx^a dx^b
\cr
 H_{(8)}&= {1\over3!} h_{abc} dx^a\wedge dx^b\wedge dx^c}
\eqn\infour
$$ 
and $e^{2\phi_{(8)}}$ depend only
 on the coordinates $\{x^a; i=1,\dots,8\}$ of the
 eight-dimensional manifold ${\cal M}_{(8)}$. Such solution
describes configurations that involve the NS-5-brane,
the KK-monopole and their various intersections and superpositions.
It remains to add a fundamental string, a pp-wave and a rotation
into this background. For this we introduce a one-form $A$,
 and two functions $g_1$ and $g_2$
on ${\cal M}_{(8)}$ which will be associated with the
 rotation, the fundamental string and pp-wave, respectively. 
Then we write the ansatz
$$
\eqalign{
ds^2&=2 g_1^{-1} dv(-du +A+g_2 dv)+ ds^2_{(8)}
\cr
H&=\lambda dv\wedge d(g_1^{-1} A)+\lambda 
du\wedge dv\wedge dg_1^{-1}+ H_{(8)}
\cr
e^{2\phi}&= g_1^{-1} e^{2\phi_{(8)}}\ ,}
\eqn\ans
$$
where $\lambda$ is a real number and $u,v$ are 
light-cone coordinates.
The functions  $g_1, g_2$, the one-form $A$
 and the constant $\lambda$
 will be determined 
by solving the Killing
spinor and field equations of supergravity. The 
metric $ds^2$ and the
three-form field strength $H$ are 
invariant under the gauge transformations
$$
\eqalign{
A&\rightarrow A+d\alpha
\cr
u&\rightarrow u+\alpha\ ,}
\eqn\bone
$$
where $\alpha=\alpha(x)$ is the parameter of 
the gauge transformations.
Our proposed solutions are therefore associated 
with a principal $U(1)$ bundle
${\cal L}$ over ${\cal M}_{(8)}$ with fibre 
coordinate $u$. The presence
of the functions $g_1$ and $g_2$ in \ans\ to
 describe a fundamental
 string and a pp-wave is standard and we shall 
not explain it further.
The addition of the gauge potential $A$ to 
describe rotation is motivated
from the way that angular momentum is computed at 
infinity. Moreover $A$
is taken to be independent from the worldvolume 
coordinates of the associated brane
since angular momentum in the context of branes in 
measured at the transverse
spatial infinity; this is a direct analogue to the
 case of black holes where
angular momentum is measured at the spatial infinity.
 The expression of the
angular momentum of a p-brane is given in 
the section below.

To find solutions to Killing spinor equations, 
it is convenient
to introduce the frame
$$
\eqalign{
e^{\underline v}&= dv
\cr
e^{\underline u}&= g_1^{-1} (-du +A+g_2 dv)
\cr
e^{\underline a}&=\tilde e^{\underline a}\ ,}
\eqn\btwo
$$
where $\tilde e$ is a frame of  $ds^2_{(8)}$. The metric and 3-form
field strength in terms of this frame are written as
$$
\eqalign{
ds^2&=2 e^{\underline v} e^{\underline u}+
\delta_{\underline a\underline b}
e^{\underline a} e^{\underline b}
\cr
H&=\lambda e^{\underline v}\wedge de^{\underline u}+ H_{(8)}\ .}
\eqn\bthree
$$
A direct computation reveals that the Killing
 spinor equations reduce to
$$
\eqalign{
(-{1\over2}\pm\lambda) g_1^{-1}\partial_bg_1 
\tilde e^j{}_{\underline a} \Gamma^{\underline
v\underline a}\epsilon_\pm&=0
\cr
({1\over2}\pm\lambda) g_1^{-1} F_{ch} 
\tilde e^c{}_{\underline a} \tilde
e^h{}_{\underline b}  \Gamma^{\underline a\underline b}\epsilon_\pm&=0
\cr
g_1^{-1}\partial_b g_2 \tilde e^b{}_{\underline a} 
\Gamma^{\underline
v\underline a}\epsilon_\pm&=0
\cr
({1\over2}\pm\lambda) g_1^{-1}\partial_bg_1 
\tilde e^b{}_{\underline a} \Gamma^{\underline
u\underline a}\epsilon_\pm&=0
\cr
D^{(\pm)}_a\epsilon_{\pm}+ {1\over2}
 ({1\over2}\pm\lambda) g_1^{-1} F_{ac}
\tilde e^c{}_{\underline b} 
\Gamma^{\underline v\underline b}&\epsilon_\pm
\cr
+{1\over4} ({1\over2}\mp\lambda)
g_1^{-1}\partial_ag_1 (\Gamma^{\underline
v\underline u}-\Gamma^{\underline
u\underline v})\epsilon_\pm&=0
\cr
\big(\Gamma^{\underline a} \tilde e^a{}_{\underline a}
 \partial_a \phi_{(8)}
\mp {1\over 3!} h_{abc} \tilde e^a{}_{\underline a}
 \tilde e^b{}_{\underline b}
\tilde e^c{}_{\underline c} 
\Gamma^{\underline a\underline b\underline c}\big)&\epsilon_\pm
\cr 
-{1\over2} \Gamma^{\underline a} \tilde e^a{}_{\underline a}
 g_1^{-1}\partial_a g_1
(1\mp \lambda [\Gamma^{\underline u \underline v} -
\Gamma^{\underline v \underline
u}])&\epsilon_\pm
\cr 
\mp {1\over2} \lambda\, g_1^{-1} F_{ab}
 \tilde e^a{}_{\underline a} \tilde e^b{}_{\underline
b} \Gamma^{\underline v}
 \Gamma^{\underline a\underline b}\epsilon_\pm&=0
  \ ,}
\eqn\killspin
$$ 
where $D^{(\pm)}_a$ are the covariant derivatives
 with torsion of ${\cal M}_{(8)}$ and
$$
F_{ab}=2 \partial_{[a} A_{b]}\ .
\eqn\bfive
$$
We
have also taken the Killing spinors
$\epsilon_\pm$ to be independent from the 
light-cone coordinates $u,v$.

To determine $g_1, g_2$ and $A$, we have also
 to use the field equations \feq. 
These are dramatically simplified provided that 
we choose $\lambda=\pm {1\over2}$.
As we shall see later for these values of
 $\lambda$ and for certain choices of
${\cal M}_{(8)}$, our solutions will preserve 
a proportion of spacetime
supersymmetry. Substituting our ansatz into the field equations for
$$
\lambda=-{1\over2}\ ,
\eqn\bsix
$$
we find
$$
\eqalign{
\partial_a \big( \sqrt{\gamma} \, e^{-2\phi_{(8)}} 
\gamma^{ab} \partial_b g_1\big)&=0
\cr
\partial_a \big( \sqrt{\gamma} \, e^{-2\phi_{(8)}} 
\gamma^{ab} \partial_b g_2\big)&=0
\cr
\partial_b\big(\sqrt{\gamma} \, e^{-2\phi_{(8)}} F^{ba}\big)-
\sqrt{\gamma}\, e^{-2\phi_{(8)}}
h^{bca} F_{bc}&=0
\ ,}
\eqn\feqgf
$$
where indices are raised and lowered using the 
metric $ds^2_{(8)}$ on ${\cal M}_{(8)}$
and $\gamma$ is the determinant of $ds^2_{(8)}$. Note that 
if we choose $\lambda=+{1\over2}$,
then the relative sign between the two terms in the
 last field equation above changes.
The two choices for $\lambda$ are symmetric, so 
without loss of generality, we shall set
$\lambda=-{1\over2}$ for the rest of the paper. In what
 follows we shall explore
the solutions of the Killing spinor
and field equations for various choices of ${\cal M}_{(8)}$ 
and investigate their
applications in strings and M-theory.

\section{Asymptotics and Angular Momentum}

Ten-dimensional spacetimes with the interpretation of p-branes are 
asymptotically
 $\bE^{(1,9)}=\bE^{(1,p)}\times \bE^{9-p}$ at the transverse
 spatial infinity,
where
$\bE^{(1,p)}$ are the worldvolume directions and $\bE^{9-p}$
 are the transverse directions of
the brane. The mass per unit volume (or tension) $M$ and the
 charge $Q$ per unit volume of a
p-brane are given as integrals over the 
$S^{8-p}\subset \bE^{9-p}$ sphere at the transverse
spatial infinity. Writing the metric as
$$
g_{MN}=\eta_{MN}+h_{MN}\ ,
\eqn\bseven
$$
where $\eta$ is the metric of ten-dimensional
 Minkowski spacetime,
the angular momentum of a p-brane per unit volume is given by
$$
 {\cJ}_{ab}= K_p\,\, {\rm lim}_{r\rightarrow \infty}\, 
\int_{S^{8-p}} \big[-x_a
\partial_ch_{0b}+x_a
\partial_th_{cb}+h_{0b}
\delta_{ac}-(a,b)\big] n^c r^{8-p} d\Omega
\eqn\bnine
$$
where $\{x^a; a=1,\dots, 9-p\}$ are the transverse
 coordinates of the brane, $r=\sqrt{
\delta_{ab} x^a x^b}$, $n^a=x^a/r$ is the outward normal 
vector of the $S^{8-p}$ sphere at
infinity, $K_p$ is a constant and $d\Omega$ is
 the volume form of $S^{8-p}$.

Our ansatz \ans\ is invariant under the action of the
Killing vector field $X=\partial_t$. Since we are
considering backgrounds that involve  either a fundamental
 string or a NS-5-brane
with possibly a wave and a rotation,  $p=1$ or $p=5$,
 respectively. In the
case of fundamental string, we take
$M_{(8)}=\bE^8$ and 
require that at the transverse spatial  infinity
$$
\eqalign{
g_1&= 1+O({1\over r^6}) 
\cr
g_2&= 0+O({1\over r^6})
\cr
A&= 0+O({1\over r^7})}
\eqn\cone
$$
and similarly for the 5-brane. In particular for the 5-brane
$$
A= 0+O({1\over r^3}) 
\eqn\ctwo
$$
The angular momentum in both
cases can be written as
$$
 {\cJ}_{ab}= K_p\, {\rm lim}_{r\rightarrow \infty} 
\int_{S^{8-p}} \big[-x_a
\partial_cA_b+A_{b}
\delta_{ac}-(a,b)\big] n^c r^{8-p} d\Omega
\eqn\ang
$$
A convenient choice for $K_p$ is 
$K_p=-{1\over (7-p) V(S^{8-p})}$, where $V(S^{8-p})$
is the volume of a (8-p)-dimensional 
sphere of radius one. 
For solutions that describe intersecting 
branes with rotation, the behaviour
at infinity is rather complicated. However we 
can compute the rotation of each brane
involved in the intersection by taking the rest 
of the branes at infinity. In
this case the expression for the angular momentum 
is again given  by \ang.

\chapter{Pure Rotation}

The geometric conditions imposed by the 
requirement of supersymmetry
and the field equations on the connection $A$
 can be most conveniently 
illustrated by taking ${\cal M}_{(8)}=\bE^8$,
 $g_1=1$ and $g_2=0$. In this case,
our ansatz \ans\ reduces to
$$
\eqalign{
ds^2&= 2dv (-du+A)+ ds^2(\bE^8)
\cr
H&=-{1\over2} dv\wedge dA\ ;}
\eqn\purerot
$$
($\lambda=-{1\over2}$). The angular momentum
of the associated spacetime is induced by $A$.

A direct substitution into Killing spinor equations 
\killspin\  reveals that
there are solutions provided that
$$
\eqalign{
\Gamma^{\underline v} \epsilon_+&=0
\cr
F_{ab} \Gamma^{ab}\epsilon_-&=0\ ,}
\eqn\killr
$$
and $\epsilon_\pm$ constant; we have made no distinction
 between coordinate and frame
indices on $\bE^8$. Moreover the field equations reduce to
$$
\partial^a F_{ab}=0\ .
\eqn\cthree
$$
Therefore $F$ satisfies the Killing spinor and field 
equations of Maxwell
theory on $\bE^8$.
  
The first Killing spinor equation can be easily solved
 giving eight Killing spinors.
To find the solutions of the second Killing spinor
 equation, we first observe that 
 the space
of two forms on
$\bE^n$ can be identified with the Lie algebra of $SO(n)$, i.e.
$$
\bigwedge^2 (\bE^n)=so(n)\ .
\eqn\cfour
$$
Then we view the Maxwell Killing spinor
equation as the equation for a spinor singlet
under an infinitesimal orthogonal rotation 
with parameter $F$.
For generic orthogonal rotations in 
$\bE^8$, there are no such singlets and
so the solution does not admit any additional
Killing spinors. So supergravity solutions 
associated with a generic
solution of the Maxwell field equations of
$A$ preserve $1/4$ of spacetime
supersymmetry. However more supersymmetry can be 
preserved by such supergravity
solutions, if $F$ takes
values in an appropriate subalgebra $h$ of
 $so(8)$. Such subalgebras are those that are 
associated with infinitesimal
orthogonal rotations which leave
certain spinors invariant (see also [\corr, \ward]).  
Examples of such subalgebras of $so(8)$ are
$h=su(2), su(3), su(4), sp(2), g_2$ and
 ${\rm spin}(7)$. In the table below
we have summarized the various fractions of 
supersymmetry preserved by the solution
for every subalgebra of $so(8)$.

\vskip 0.3cm
$$
\vbox{\settabs 8\columns 
\+ {\rm Algebra}& $so(8)$ & ${\rm spin}(7)$&$g_2$ 
& $sp(2)$ & $su(4)$ & $su(3)$ & $su(2)$
 \cr
\+{\rm Susy} &${1\over 4}$& ${9\over32}$&${10\over32}$
& ${11\over32}$ & ${10\over32}$&
$~{3\over8}$&$~{1\over2}$\cr}
 $$
\vskip 0.2cm

For the computation of these fractions, we have first
 decomposed the sixteen-dimensional
 representations ${\bf 16}_s$ and ${\bf 16}_c$ of 
${\rm Cliff}(\bE^{(1,9)})$ under ${\rm
Spin}(1,1)\times {\rm Spin}(8)$ as
$$
\eqalign{
{\bf 16}_s&\rightarrow ({\bf1}_s, {\bf 8}_s)
\oplus ({\bf1}_c, {\bf 8}_c)
\cr
{\bf 16}_c&\rightarrow ({\bf1}_s, {\bf 8}_c)
\oplus ({\bf1}_c, {\bf 8}_s)\ .}
\eqn\dec
$$
The solutions of the  first Killing spinor equation  
in \killr\ are either in the
subspace
$({\bf1}_s, {\bf 8}_s)$ or in the subspace
$ ({\bf1}_s, {\bf 8}_c)$ of ${\bf 16}_s$
depending on whether $\Gamma^{\underline v}\epsilon_+=0$ is 
related to a anti-chiral or to a
chiral projection of ${\rm Cliff}(1,1)$, respectively. 
Without loss of generality we can
choose the former case, so the Killing spinors are
 in $({\bf1}_s, {\bf 8}_s)$ subspace of
${\bf 16}_s$. The remaining Killing spinors arise 
as  singlets in the
decomposition of  ${\bf 8}_s$ and ${\bf 8}_c$ that 
appears in the second equation of \dec\
under the associated subalgebra  $h$ of $so(8)$ in 
the table above. Similar
computations have been done elsewhere (see for example [\gtpap]) and we 
shall not give more details here.

\section{$h=su(2), su(3)$ and $su(4)$}

We begin the construction of explicit
 supergravity solutions by first taking
$F$ to take values in $h=su(n)$ for $n=2,3,4$. 
Such connections have support in a subspace
$\bE^{2n}$ of $\bE^8$ and they are characterized
 by a complex structure $K$ on $\bE^{2n}$
which determines the embedding of $SU(n)$ in 
$SO(2n)\subset SO(8)$. For $h=su(2)$,
$A$ is either self-dual or anti-self-dual 
connection depending on the choice
of complex structure $K$   relative to the 
choice of orientation of  $\bE^4$.
For $h=su(n)$, $n=3,4$, $A$ is a Hermitian-Einstein
 connection on $\bE^6$ and $\bE^8$,
respectively [\donaldson]. To continue, we choose complex coordinates 
$\{z^\alpha; \alpha=1, \dots, n\}$
with respect to $K$. The conditions on $F$ for $h=su(n)$ are
$$
\eqalign{
F_{\alpha\beta}&=0
\cr
\delta^{\alpha\bar\beta} F_{\alpha\bar\beta}&=0\ .}
\eqn\hereist
$$
Note that the flat metric is hermitian with 
respect to $K$ and that the above
conditions on $F$ imply the Maxwell field 
equations using the Jacobi identity.

To find the solutions of the above
 equations, we first observe that
the first one implies that
$$
\eqalign{
A_\alpha&=\partial_\alpha U
\cr
A_{\bar \alpha}&=\partial_{\bar \alpha} \bar U\ ,}
\eqn\done
$$
where $U$ is a complex function.
Substituting this into the second equation 
in \hereist, we find that
$$
\delta^{\alpha\bar\beta} 
\partial_\alpha \partial_{\bar\beta} \big(U-\bar U\big)=0\ .
\eqn\dtwo
$$   
This equation can be easily solved to find that
$$
U-\bar U=i f
\eqn\dthree
$$
where
$$
f=1+\sum^N_i {\mu_i \over |z-z_i|^{2n-2}}
\eqn\harm
$$
is a real harmonic function on $\bR^{2n}$.
The above equations do not determine the real part of 
$U$ but this is expected
since it is a gauge degree of freedom. So up 
to a gauge transformation,
the solution is
$$
\eqalign{
A_\alpha&=i \partial_\alpha f
\cr
A_{\bar \alpha}&=-i \partial_{\bar \alpha}f\ ,}
\eqn\dfour
$$
which can be  rewritten in real coordinates  as
$$
A_a= K^b{}_a \partial_b f\ .
\eqn\dfive
$$
In four dimensions, the solution in this form 
has already been given in [\popov, \gibpap]. 
Moreover it turns out that
if the K\"ahler form of
$K$ is self-dual, then $A$ is anti-self-dual
and vice-versa. In fact in this case the 
solution can be generalized.
For this we use that $h=su(2)=sp(1)$ and 
observe the Hermitian-Einstein
condition on $F$ can be imposed with any of 
the three complex structures,
or a linear combination of them, that define 
the embedding of $sp(1)$ in $so(4)$.
We remark that independently from the choice of
complex structure, $F$ is 
either a
self-dual or anti-self-dual two-form.
Let $\{I_r; r=1,2,3\}$ be the quaternionic 
structure that defines
 the embedding of $sp(1)$ in
$so(4)$. Using the linearity of the equations and 
the remark above, 
the most general solution
can be written as
$$
A_a=\sum^N_i \sum_{r=1}^3(I_r)^b{}_a 
\partial_b{\mu^r_i \over |x-x_i|^{2}}\ .
\eqn\mone
$$

We shall see that for the investigation 
the asymptotic behaviour
of  supergravity solutions, it is sufficient to
consider
spherically symmetric rotations. The rest of the 
solutions can be thought as  \lq
parallel' superpositions of the spherically 
symmetric ones. For spherically
symmetric solutions,  we have
$$
A_a= K^b{}_a \partial_b {\mu \over |x|^{2n-2}}
\eqn\mtwo
$$
for $n=3$ and $n=4$, and
$$
A_a= \sum_{r=1}^3(I_r)^b{}_a \partial_b{\mu^r \over |x|^{2}}
\eqn\mthree
$$
for $n=2$, respectively. 
The solutions for $n=2$ and $n=4$ can be used  to add rotation 
to a NS-5-brane
and  to a fundamental string, respectively. The 
case $n=3$ can be used to add rotation
to a D-3-brane or a compactified fundamental 
string. The latter case  will lead
a rotating black hole in seven dimensions.
Using the expression for the angular momentum \ang, we find that
$$
{\cJ}_{ab}= \mu K_{ab}
\eqn\mfour
$$ 
for $n=3,4$, and
$$
{\cJ}_{ab}=\sum_{r=1}^3 \mu^r (I_r)_{ab}
\eqn\mfive
$$
for $n=2$. In both cases, the angular momentum 
has one independent eigenvalue.

\section{$h=sp(2)$}

Let $\{{\bf K}_r; r=1,2,3\}$ be a quaternionic 
structure in $\bE^8$
that determines the embedding of $Sp(2)$ in $SO(8)$. 
The condition that
$F$ is in $sp(2)$ is equivalent to the 
condition that $F$ is a (1,1)
two-form with respect all three complex 
structures  $\{{\bf K}_r; r=1,2,3\}$,
i.e.
$$
F_{ab}({\bf K}_r)^a{}_c ({\bf K}_r)^b{}_d=F_{cd}
\eqn\msix
$$
(no summation over $r$). To find the conditions on 
$A$, it is convenient to introduce
a coordinates  $\{x^{i\mu}; i=1,2; \mu=1, \dots, 4\}$ 
on $\bE^8$ such that
$$
{\bf K}_r{}^{i\mu}{}_{j\nu}=\delta^i{}_j (K_r)^\mu{}_\nu
\eqn\mseven
$$
where $\{K_r; r=1,2,3\}$ is a quaternionic structure is $\bE^4$.
Then since $F$ is (1,1)-form with respect to all 
three complex structures  $\{{\bf K}_r;
r=1,2,3\}$, we choose the first one and can write
$$
A_{i\mu}= (K_1)^\nu{}_\mu \partial_{i\nu}f\ ,
\eqn\hermiti
$$
where $H$ is a real function, as in the 
case of Hermitian-Einstein 
connections in the previous
section. This leads to an $F$ which is 
(1,1)-form with respect to ${\bf K}_1$.
It is straightforward to show that for $F$ 
to be (1,1)-form with respect
to the remaining complex structures then
$$
\eqalign{
\delta^{\mu\nu}\partial_{i\mu} \partial_{j\nu} f&=0 
\cr
\partial_{i\mu}\partial_{j\nu}f&=\partial_{j\mu}\partial_{i\nu}f}
\eqn\meight
$$
Solving these equations\foot{For the 
non-abelian case see [\kent].}, we find that
$$
A_{i\mu}=\sum_{\{(p, a)\}}\sum_{r=1}^3 
(K_r)^\nu{}_\mu \partial_{i\nu} {\mu^r\big((p, a)\big)
\over |p_i x^{i\mu}-a^\mu|^2}\ ,
\eqn\sptwosol
$$
where the parameters $\{p_i, i=1,2; a^\mu,
\mu=1,\dots,4\}$ of the solution are the real 
numbers. To write this solution, we have
also used the fact that
$A$ can be written in three different but 
equivalent ways as \hermiti\ with respect
to each complex structure $\{K_r; r=1,2,3\}$.

The interpretation of the solutions  
\sptwosol\ is as a superposition of
four-dimensional abelian instantons 
(or anti-instantons) at $Sp(2)$ angles.
This follows from the analysis in [\gpandrew]
 of the non-abelian case. Therefore
the abelian gauge potentials \sptwosol\ can 
be used to rotate intersecting brane
configurations. This is achieved by using 
each four-dimensional abelian instanton involved
in the superposition to rotate a brane 
involved in the intersection.  We
shall verify this when we investigate the rotation of
 intersecting 5-branes at angles.  If we
take all the abelian instantons at infinity apart from one 
and compute the angular momentum
of the associated spacetime,
we shall find that
$$
(\cJ)_{ab}= \sum_{r=1}^3 \mu^r (K_r)_{\mu\nu}
\eqn\mnine
$$
which is that of the $n=2$ case of the previous section.

We finally remark that the condition that $F$ 
is in $sp(2)$ can also be written
as self-duality like condition
$$
{1\over2}\Omega_{ab}{}^{cd} F_{cd}=F_{ab}\ ,
\eqn\mten
$$
where
$$
\Omega=\sum_{r=1}^3 {\bf \omega}_r\wedge{\bf \omega}_r
\eqn\none
$$
where $\omega_r$ is the K\"ahler form of ${\bf K}_r$.

\section{$h=spin(7)$ and $g_2$}

We shall first give the solutions for which $F$ is in $spin(7)$.
For this, let $\varphi$ be the 3-form 
with components the structure constants
of imaginary unit octonions and $\phi$ be its dual in $\bE^7$, i.e.
$$
\phi_{ijk\ell}={1\over6} \epsilon_{ijk\ell}{}^{pqs} \varphi_{pqs}\ ,
\eqn\ntwo
$$
where $i,j,k, \ell, p,q,s=1,\dots,7$.
Then a $spin(7)$-invariant self-dual 4-form 
$\omega$ in $\bE^8$ can
be defined as
$$
\eqalign{
\omega_{ijk8}&=\varphi_{ijk}
\cr
\omega_{ijk\ell}&=\phi_{ijk\ell}\ .}
\eqn\mnthree
$$
For later use, we remark that the four-form 
$\omega$ satisfies
$$
\omega^{abch }\omega_{defh}=6 
\delta^{abc}_{def}- 9\delta^{[a}_{[d} \omega^{bc]}{}_{ef]}
\eqn\mnfour
$$
where $a,b,c,d,e,f,h=1,\dots,8$.
We introduce the orthogonal projectors 
$$
\eqalign{
(P_1)^{ab}{}_{cd}&= {3\over4} 
(\delta^{ab}{}_{cd}+{1\over6} \omega^{ab}{}_{cd})
\cr
(P_2)^{ab}{}_{cd}&= {1\over4} 
(\delta^{ab}{}_{cd}-{1\over2} \omega^{ab}{}_{cd})}
\eqn\mnsix
$$
on $\bigwedge^2(\bE^8)$. 
These projectors are associated with an orthogonal 
decomposition of $\bigwedge^2(\bE^8)$ into a 
21-dimensional subspace [\fn], which can be identified
 with $spin(7)$,  and a seven-dimensional
subspace, respectively.
The condition that $F$ is in $spin(7)$ is 
therefore $P_2 F=0$ or equivalently
$$
{1\over2}\omega_{ab}{}^{cd} F_{cd}=F_{ab}\ .
\eqn\mmone
$$
To find solutions to this equation, we introduce the ansatz
$$
A_a=v^i (I_i)^b{}_a \partial_b f
\eqn\mmtwo
$$
where
$$
\eqalign{
(I_i)^8{}_j&=\delta_{ij}
\cr
(I_i)^j{}_8&=-\delta_i^j
\cr
(I_i)^k{}_j&=\varphi_i{}^k{}_j}
\eqn\mmthree
$$
and $v$ is a (constant) vector in $\bE^7$. The 
matrices $\{I_i; i=1,\dots,7\}$ satisfy
$I_i I_j+I_jI_i=-2\delta_{ij}$ and so they give  
an irreducible representation 
of ${\rm Cliff}(\bR^7)$ equipped
with the negative definite inner product.
Using the identities
$$
\eqalign{
\phi^{mijk} \varphi_{mpq}&=-6 \delta^{[i}{}_{[p}
\varphi^{jk]}{}_{q]}
\cr
\varphi^{mij}\varphi_{mk\ell}&=
2\delta^{ij}{}_{k\ell}- \phi^{ij}{}_{k\ell}\ ,}
\eqn\mmfour
$$
a direct substitution of the ansatz into 
the condition $P_2 F=0$ reveals 
that
$$
\delta^{ab}\partial_a\partial_b f=0\ ,
\eqn\mmfive
$$
for any choice of $v$. So  $f$ is a 
harmonic function in $\bE^8$.
Therefore for this class of  solutions
$$
A_a=\sum_{n=1}^N \sum_{i=1}^7 (I_i)^b{}_a 
\partial_b{\mu_n^i\over |x-x_n|^6}\ .
\eqn\spinrot
$$
It is clear that the above abelian $U(1)$ 
gauge fields can be used to rotate
fundamental strings and M-2-branes. The 
spherically symmetric solution is
$$
A_a= \sum_{i=1}^7 (I_i)^b{}_a \partial_b{\mu^i\over |x|^6}\ .
\eqn\mfone
$$
The angular momentum of the spacetime 
associated with this solution is
$$
({\cJ})_{ab}= \sum_{i=1}^7 \mu^i (I_i)_{ab}\ .
\eqn\mftwo
$$

Next we shall examine the case where 
$F$ is in $g_2$. We again 
introduce the projectors
$$
\eqalign{
(P_1)^{ij}_{k\ell}&= {2\over3}
 \big(\delta^{ij}{}_{k\ell} +{1\over4} 
\phi^{ij}{}_{k\ell}\big)
\cr
(P_2)^{ij}_{k\ell}&= {1\over3} 
\big(\delta^{ij}{}_{k\ell} -{1\over2}
\phi^{ij}{}_{k\ell}\big)}
\eqn\mgone
$$
which introduce an orthogonal decomposition 
of $\bigwedge^2(\bR^7)$ into a 14-dimensional
subspace, that can be identified with $g_2$, 
and a seven-dimensional subspace. Since $P_2$ 
projects onto the seven-dimensional subspace, 
the condition that $F$ is in $g_2$ is
$P_2(F)=0$ or equivalently
$$
{1\over2}\omega_{ij}{}^{k\ell} F_{k\ell}=F_{ij}\ .
\eqn\gtwocon
$$
To solve this equation, we write the ansatz
$$
A_i= v^k \varphi_k{}^j{}_i \partial_j f
\eqn\wone
$$
in analogy with the $spin(7)$ case above. Then for 
$F$ to be in $g_2$, we find that
$$
\varphi_{ij}{}^k v^\ell 
\partial_k\partial_\ell f-\varphi_{ijk} v^k
\delta^{m\ell}\partial_m\partial_\ell f=0\ .
\eqn\wtwo
$$
Since the above equation is linear $v$, we can 
take with loss of generality
$v$  to have length one. To solve the above 
equation let us assume that
$v=(0,\dots,0,1)$. Now if we take $f$ to be 
independent from $x^7$, then the
above equation implies that $f$ is 
harmonic on the hyperplane in $\bE^7$
orthogonal to $v$. So for this choice of 
$v$ a typical solution is
$$
f=1+\sum_{n=1}^N{\mu_n\over |\tilde x-\tilde x_n|^4}
\eqn\wthree
$$
where $\tilde x$ are the first six coordinates 
of $\bE^7$. For the most general
solution of this type using the linearity of 
the condition \gtwocon\ in $A$,
we can sum over different choices of $v$.  This 
leads to the following solution for $A$:
$$
A_i=\sum_{\{v\}} \sum_{n=1}^N  v^k 
\varphi_k{}^j{}_i \partial_j
{\mu_n\over |x_v-(x_v)_n|_v^4}\ ,
\eqn\gtworot
$$
where $x_v$ are the coordinates of the
 hyperplane $P_v$ orthogonal to $v$ and 
$|\cdot|_v$ is the norm on  $P_v$ 
 induced by the standard norm on $\bE^7$.

The solutions of the condition that $F$ is in $g_2$
 has many similarities with the 
solutions we have found   for the
condition that $F$ is in $sp(2)$ in 
section (3.2). In the latter case,  the
expression for $A$ is given by summing over 
 four-dimensional subspaces of $\bE^8$ while in the former 
the expression for $A$
is given by summing over hyperplanes
 in $\bE^7$. Therefore
as in the $sp(2)$ case, the abelian 
$U(1)$ gauge field
can be interpreted as a superposition 
of abelian Hermitian-Einstein
instantons for which $F$ is in $su(3)$ 
along hyper-planes in $\bE^7$.
This interpretation is consistent with 
the power law decay of the gauge
field at infinity. 

The $U(1)$ gauge fields \gtworot\ can be 
used to rotate either intersecting
brane configurations or appropriately 
compactified fundamental strings
and M-2-branes. The rotation of the 
spacetime associated with 
the spherically symmetric
solution
$$
A_i=   v^k \varphi_k{}^j{}_i \partial_j
{\mu\over |x_v|_v^4}\ ,
\eqn\sgtworot
$$
is
$$
({\cJ})_{ij}=\mu  v^k \varphi_{kij}\ ,
\eqn\wfour
$$
where we have performed the integration at 
the $S^5$ sphere at infinity of the 
hyperplane $P_v$.

\section{$h=so(8)$}

There are many solutions to the Maxwell 
equations that include for example
elecromagnetic waves. Here we shall seek solutions 
for which $F$ spans other
subalgebras of
$so(8)$ from those that we have mentioned in 
the previous sections.
Such configurations as solutions of the Maxwell 
theory are not supersymmetric.
Moreover, we shall require that they resemble the
 supersymmetric solutions
that we have constructed in the previous sections.  An example of 
this is to  consider a
slight generalization of the Hermitian-Einstein 
condition on $F$ by letting
$F$ to take values  in $u(n)$ instead of the
$su(n)$ subalgebra of $so(2n)$, i.e.
$$
F_{ab}=\Lambda (\omega_K)_{ab}\ .
\eqn\ghen
$$
The Maxwell field equations imply 
that $\Lambda$ is constant.
Again $F$ is a (1,1)-form with respect 
to the complex structure $K$ and
a straightforward computation reveals 
that a solution of \ghen\ is
$$
A_a=K^b{}_a \partial_b 
f-{1\over2}\Lambda (\omega_K)_{ab} x^b\ ,
\eqn\lherein
$$
where $f$ is that of \harm. The advantage 
of considering \ghen\ is that
it admits an obvious generalization on
Hermitian manifolds which
include the Calabi-Yau and hyper-K\"ahler manifolds. 
The disadvantage
of such solution is that the second 
term in \lherein\ may lead to spacetimes
that are not asymptotically flat.

An alternative way to construct 
non-supersymmetric  solutions is
by superposing the supersymmetric 
solutions found in the previous 
sections using the
linearity of Maxwell equations. To 
demonstrate this, we shall first
construct a solution for which $F$ 
takes values in $so(4)$ by superposing
and instanton and anti-instanton solution. For this we
 introduce two commuting 
quaternionic structures $\{I_r; r=1,2,3\}$ and
$\{J_r; r=1,2,3\}$ on $\bR^4$ 
associated with the decomposition $so(4)=su(2)\oplus
su(2)$. Then a solution  with $F$ in $so(4)$ is
$$
A_a=\sum^N_{i=1} \sum_{r=1}^3(I_r)^b{}_a 
\partial_b{\mu^r_i \over
|x-x_i|^{2}}+\sum^{N'}_{i=1}
\sum_{r=1}^3(J_r)^b{}_a 
\partial_b{\tilde \mu^r_i \over |x-x_i|^{2}}\ .
\eqn\sofourrot
$$
These abelian $U(1)$ gauge 
field can be used to rotate  
NS-5-branes\foot{To find such solutions 
the ansatz \ans\ that we are using
has to be modified.}
and compactified strings  which
will lead to rotating black holes in five-dimensions. 
A spherically symmetric $U(1)$ gauge field is
$$
A_a=\sum_{r=1}^3\big[\mu^r 
(I_r)^b{}_a + \tilde \mu^r
(J_r)^b{}_a \big] \partial_b{1\over |x|^{2}}
\eqn\wsix
$$
and the angular momentum of the associated spacetime is
$$
({\cJ})_{ab}=\sum_{r=1}^3\big[ \mu^r 
(I_r)_{ab}+\tilde \mu^r (J_r)_{ab}\big]\ .
\eqn\wseven
$$
Observe that the angular momentum is a vector in $so(4)$ and it 
has two independent
eigenvalues equal to the rank of $so(4)$. 

The above procedure can be easily 
generalize to superpose Hermitian-Einstein connections
$A$ to construct solutions for which $F$ 
is in $so(8)$. For this we consider
complex structures $\{K_q; q=1, \dots, \ell\}$ 
on $\bE^8$ and superpose the
Hermitian-Einstein connections with 
respect to each $K_q$. A solution is
$$
A_a=\sum_{q=1}^\ell (K_q)^b{}_a \partial_b f_q
\eqn\soeightrot
$$
where $\{f_q; q=1, \dots, \ell\}$ are distinct harmonic 
functions similar to those in \harm\ for $n=4$. Then
 $F$ spans the subspace $\cup_{q=1}^\ell
su(4)\subset so(8)$, where each $su(4)$ is associated 
with a complex structure $\{K_q; q=1,\dots, \ell\}$. There is
always a choice of complex structures
$\{K_q; q=1,
\dots,
\ell\}$ such that
$F$ spans $so(8)$.  The $U(1)$ gauge field 
\soeightrot\ can be used
to rotate a fundamental string or a M-2-brane. 
A spherically symmetric solution is
$$
A_a= \sum_{q=1}^\ell (K_q)^b{}_a \partial_b {\mu_q\over |x|^6}
\eqn\weighta
$$
and the angular momentum of the associated spacetime is
$$
({\cJ})_{ab}= \sum_{q=1}^\ell \mu_q (K_q)_{ab}\ .
\eqn\wnine
$$
In general, ${\cJ}$ would have four independent eigenvalues
equal to the rank of $so(8)$.

We can also do  similar superpositions using 
connections for which $F$ is in $sp(2)$.
For this consider for example the 
two quaternionic structures on $\bE^8$,
$$
\eqalign{
{\bf I}_r{}^{i\mu}{}_{j\nu}&=
\delta^i{}_j (I_r)^\mu{}_\nu
\cr
{\bf J}_r{}^{i\mu}{}_{j\nu}&=
\delta^i{}_j (J_r)^\mu{}_\nu\ ,}
\eqn\bigqst
$$
where $\{I_r; r=1,2,3\}$ and $\{J_r; r=1,2,3\}$ 
are those of the $so(4)$ case above.
Then the curvature $F$ of the connection
$$
A_{i\mu}=\sum_{\{p_i, a\}}\sum_{r=1}^3 
(I_r)^\nu{}_\mu \partial_{i\nu} {\mu^r(p, a) \over |p_i
x^{i\mu}-a^\mu|^2}+\sum_{\{\tilde p_i, \tilde a\}}
\sum_{r=1}^3 (J_r)^\nu{}_\mu \partial_{i\nu}
{\tilde\mu^r(\tilde p, \tilde a)
\over |\tilde p_i x^{i\mu}-\tilde a^\mu|^2}\ ,
\eqn\wten
$$
spans the subspace $sp(2)\cup sp(2)\subset so(8)$, 
where each $sp(2)$ is associated
with the quaternionic structures \bigqst\ above. 
It is clear that there are many
other ways of constructing new solutions for example by
 superposing the different methods
proposed above. The angular momentum
 induced on branes or intersecting
branes by these $U(1)$ gauge fields can be easily computed
and we shall not present the result here. 
The above results can be
easily extended to any suitable dimension.

\section{Strings, Waves and Rotations}

It is straightforward to superpose the
rotated spacetimes we have
constructed with a pp-wave and a fundamental 
string for ${\cal M}_{(8)}=\bE^8$.
The field equations simply imply that the 
functions $g_1$ and $g_2$ associated
with the string and the wave are harmonic 
functions on $\bE^8$.
Moreover, the field equations for $A$ 
do not alter with the
addition of the string and the wave and so 
the expressions that we have given for $A$
in the previous sections are still valid.
The supersymmetry preserved by such solutions 
depends on the type
of rotation and whether the configuration 
involves a wave or a string or both.
Since the supersymmetry of these 
configurations without  rotation
has been investigated in the literature, 
here we shall consider three cases
of solutions with (i) rotation and wave, 
(ii) rotation and string and
(iii) rotation, wave and string. We shall 
do the analysis without reference
to a particular manifold ${\cal M}_{(8)}$ and only at the end 
we shall specialize to  ${\cal M}_{(8)}=\bE^8$. We 
shall also choose $\lambda=-{1\over2}$.

$(i)\,$ For such configurations, we allow $F, g_2$ 
to depend on $x\in \cMe$ and set
$g_1=1$. The third Killing spinor equation 
in \killspin\ implies that 
$$
\Gamma^{\underline v}\epsilon_{\pm}=0
\eqn\killrwa
$$
and the remaining reduce to
$$
\eqalign{
F_{ch} \tilde e^c{}_{\underline a}\tilde e^h{}_{\underline b}
\Gamma^{\underline a\underline
b} \epsilon_-&=0
\cr
D_a^{(\pm)}\epsilon_\pm=0
\cr
\big(\Gamma^{\underline a} \tilde e^a{}_{\underline a}
\partial_a\phi_{(8)}\mp
{1\over3!} h_{abc} \tilde e^a{}_{\underline a}
 \tilde e^b{}_{\underline b}
\tilde e^c{}_{\underline c} \Gamma^{\underline a\underline
b\underline c}\big)\epsilon_\pm&=0\ .}
\eqn\killrwb
$$
So for ${\cal M}_{(8)}=\bE^8$, these imply 
that $\epsilon_\pm$ are constant which 
satisfy \killrwa\ and in addition $\epsilon_-$ 
satisfies the second equation
in \killr, i.e. $F_{ab} \Gamma^{ab} \epsilon_-=0$. 
Using the decomposition \dec, the
solutions of
condition \killrwa\  lie either in the subspace
$({\bf1}_s, {\bf 8}_s)\oplus ({\bf1}_s, {\bf 8}_c)$ or 
in the subspace 
 $({\bf1}_c, {\bf 8}_c)
\oplus ({\bf1}_c, {\bf 8}_s)$ of 
 ${\bf 16}_s\oplus {\bf 16}_c$.
The fraction of supersymmetry preserved 
depends on the way we embed the
subalgebra $h$ that $F$ lies in $so(8)$. 
If we use the embeddings
of [\salamon], then the fractions of supersymmetry 
preserved  for spinors in the  subspaces 
$({\bf1}_s, {\bf 8}_s)\oplus ({\bf1}_s, 
{\bf 8}_c)$ and $({\bf1}_c, {\bf 8}_c)
\oplus ({\bf1}_c, {\bf 8}_s)$ are 
summarized in the table below, respectively.

\vskip 0.3cm
$$
\vbox{\settabs 8\columns 
\+ {\rm Algebra}& $so(8)$ & ${\rm spin}(7)$&$g_2$ 
& $sp(2)$ & $su(4)$ & $su(3)$ & $su(2)$
 \cr
\+{\rm Susy} &${1\over 4}$& ${1\over 4}$&${1\over 4}$
& ${1\over 4}$ & ${1\over4}$&
$~{10\over32}$&$~{3\over8}$\cr
\+{\rm Susy} &${1\over 4}$& ${9\over32}$&${10\over32}$
& ${11\over32}$ & ${10\over32}$&
$~{10\over32}$&$~{3\over8}$\cr}
 $$
\vskip 0.2cm

$(ii)\,$ For such configurations, we allow $F, g_1$ 
to depend on $x\in \cMe$ and set
$g_2=0$. The first and fourth Killing spinor 
equations in \killspin\ imply that
$$
\eqalign{
\Gamma^{\underline v}\epsilon_+&=0
\cr
\Gamma^{\underline u}\epsilon_-&=0\ ,}
\eqn\killrsa
$$
respectively. To continue, we set\foot{ The 
expression is asymmetric
in $\epsilon_+$ and $\epsilon_-$ which
 is  not conventional. However
it is due to the choice of the frame for 
our metric. One can find a 
symmetric expression
using a frame rotation.}
$$
\eqalign{
\epsilon_+&= g_1^{-{1\over2}}\eta_+
\cr
\epsilon_-&=\eta_-}
\eqn\killsub
$$
Then, using $\Gamma^{\underline u\underline v}+
\Gamma^{\underline v\underline u}=2$,
the rest of the Killing spinor equations in \killspin\ reduce to
$$
\eqalign{
g_1^{-1}F_{ch} \tilde e^c{}_{\underline a}
\tilde e^h{}_{\underline b}
\Gamma^{\underline a\underline
b} \eta_-&=0
\cr
D_a^{(\pm)}\eta_\pm=0
\cr
\big(\Gamma^{\underline a} 
\tilde e^a{}_{\underline a}\partial_a\phi_{(8)}\mp
{1\over3!} h_{abc} \tilde e^a{}_{\underline a} 
\tilde e^b{}_{\underline b}
\tilde e^c{}_{\underline c} 
\Gamma^{\underline a\underline
b\underline c}\big)\eta_\pm&=0\ .}
\eqn\killrsb
$$
So for ${\cal M}_{(8)}=\bE^8$, these imply that 
$\eta_\pm$ are constant which 
satisfy \killrsa\ and in addition 
$\epsilon_-$ satisfies 
$F_{ab} \Gamma^{ab} \epsilon_-=0$.
Using the decomposition \dec, 
the solutions of the condition 
\killrsa\
lie  either in the subspace 
$({\bf1}_s, {\bf 8}_s)\oplus ({\bf1}_c, {\bf 8}_s)$
 or in the
subspace 
 $({\bf1}_c, {\bf 8}_c)
\oplus ({\bf1}_s, {\bf 8}_c)$ of 
${\bf 16}_s\oplus {\bf 16}_c$. Choosing 
the embedding
of $h$ in $so(8)$ as in case (i), the 
fractions of supersymmetry preserved in each
of the above two cases are summarized 
in the table below, respectively.

\vskip 0.3cm
$$
\vbox{\settabs 8\columns 
\+ {\rm Algebra}& $so(8)$ & ${\rm spin}(7)$&$g_2$ & 
$sp(2)$ & $su(4)$ & $su(3)$ & $su(2)$
 \cr
\+{\rm Susy} &${1\over 4}$& ${9\over32}$&${10\over32}$& 
${11\over32}$ & ${10\over32}$&
$~{10\over32}$&$~{3\over8}$\cr
\+{\rm Susy} &${1\over 4}$& ${1\over 4}$&${1\over 4}$& 
${1\over 4}$ & ${1\over4}$&
$~{10\over32}$&$~{3\over8}$\cr}
 $$
\vskip 0.2cm

$(iii)\,$ For solutions with rotation, 
fundamental string and wave, the third and 
fourth Killing spinor equations of \killspin\ imply that
$$
\eqalign{
\Gamma^{\underline v} \epsilon_+&=0
\cr
\epsilon_-&=0\ .}
\eqn\killrswa
$$
the rest of the Killing spinor equations reduce to
$$
\eqalign{
D_a^{(+)}\eta_+=0
\cr
\big(\Gamma^{\underline a} 
\tilde e^a{}_{\underline a}\partial_a\phi_{(8)}-
{1\over3!} h_{abc} 
\tilde e^a{}_{\underline a} \tilde e^b{}_{\underline b}
\tilde e^c{}_{\underline c} \Gamma^{\underline a\underline
b\underline c}\big)\eta_+&=0\ ,}
\eqn\killrswb
$$
where we have set $\epsilon_+=g_1^{-{1\over2}}\eta_+$. 
So for ${\cal M}_{(8)}=\bE^8$,
these imply that $\eta_+$ is constant 
which satisfies \killrswa. Such solution
preserves 1/4 of supersymmetry 
for any choice of solution for $A$.

\chapter{Rotations in toric hyper-K\"ahler Spaces}

Toric hyper-K\"ahler eight-dimensional spaces 
admit two tri-holomorphic
commuting Killing vector fields. To describe 
the metric of
such space, we introduce the coordinates  
$\{(\tau_i, y^{ir}); i=1,2; r=1,2,3\}$
on $\cMe$, where $\tau_i$ are the coordinates 
adopted along the Killing
vector fields. Then the metric [\poon, \gtpap] is
$$
ds^2_{(8)}=U^{ij} (d\tau_i+\omega_i) 
(d\tau_j+\omega_j) +U_{ij}\delta_{rs} dy^{ir}
dy^{js}\ ,
\eqn\torhyp
$$
where $\omega_i=\omega_{i, jr}dy^{jr}$,  
$U^{ij} U_{jk}=\delta^i{}_k$ and
$$
\eqalign{
\partial_{ir}U_{jk}&=\partial_{jr}U_{ik}
\cr
\delta^{tu}\epsilon_{trs} \partial_{iu}U_{jk}&= 
\partial_{[ir} \omega_{|k|, jr]}\ .}
\eqn\weleven
$$
One  can find that
$$
U_{ij}=U^\infty_{ij}+\sum_{\{(p, a)\}}\mu\big((p, a)\big)
{p_i p_j\over |p_iy^i-a|}\ ,
\eqn\kone
$$
where $\{U^\infty_{ij}\}$ is suitably chosen  
constant matrix and $\{(p_i, a^r); i=1,2;
r=1,2,3\}$ are the parameters of the solution. The 
interpretation of this solution
in the context of strings or M-theory is as 
superposition of KK-monopoles at $Sp(2)$
angles.
We can easily add a wave and a string in the
configurations in which case
$g_1$ and $g_2$ will be harmonic functions 
with respect to the \torhyp. This
equation is considerably simplified if we 
take $g_1$ and $g_2$ to be invariant
under the Killing isometries. In this case, $g_1$ satisfies
$$
U^{ij} \delta^{rs} \partial_{ir}\partial_{js} g_1=0
\eqn\ktwo
$$
and similarly for $g_2$.
A solution of this equation is
$$
g_1=1+\sum_{\{(p, a)\}}\mu\big((p, a)\big){1\over |p_iy^i-a|}\ ,
\eqn\kthree
$$
and similarly for $g_2$. The parameters 
$\{(p, a)\}$ of the metric, $g_1$ and $g_2$ may be
different. But if we require for a wave and for
a string to lie on each KK-monopole involved
in the superposition, the we must take all of them to be the same
up to a scale.

It remains to solve the equation for the 
rotation. In this case there are two
natural Maxwell field strengths  with 
values in $sp(2)$. The gauge
 potentials are the duals
of the Killing vector fields of the 
toric hyper-K\"ahler metric, i.e.
$$
A^{(i)}= =U^{ij}  (d\tau_j+\omega_j)\ .
\eqn\kfour
$$
The associated two-form field strengths 
$F^{(i)}=dA^{(i)}$ are in $sp(2)$ because
the Killing vectors fields are tri-holomorphic. 
Moreover $F^{(i)}$, $i=1,2$, solve the Maxwell
field equations either as consequence of the 
Jacobi identity together with the
condition that they lie in $sp(2)$ or 
equivalently because the toric hyper-K\"ahler
spaces are Ricci flat [\gibbons]. The most 
general $A$ that can substitute in our
ansatz is a linear combination
$$
A=c_i A^{(i)}=c_i U^{ij}  (d\tau_j+\omega_j)
\eqn\kfive
$$
of the two $U(1)$ gauge potentials above, where 
$\{c_i, i=1,2\}$ is a constant vector.

\chapter{Rotations and Intersecting Branes}

There are three different orthogonal 
intersections of  NS-5-branes. These are 
(i) any two NS-5-branes intersecting on a 
3-brane, (ii) any
three NS-5-branes intersecting  on a 
string and (iii) any two NS-5-branes
intersecting at a string. The latter 
intersection can be generalized
to a multiple NS-5-brane intersection 
at $Sp(2)$ angles.
All these intersections are associated 
with a $\cMe$ geometry. In what
follows, we shall be mainly concerned 
with $Sp(2)$-intersecting NS-5-branes
 on a string.
To  describe the eight-dimensional geometry 
that arise in this case,
we introduce coordinates 
$\{x^{i\mu}; i=1,2; \mu=1,\dots,4\}$ on $\cMe$
and two quaternionic structures
$$
\eqalign{
{\bf J}_r{}^{i\mu}{}_{j\nu}&=\delta^i{}_j J_r{}^\mu{}_\nu
\cr
{\bf I}_r{}^{i\mu}{}_{j\nu}&=\delta^i{}_j I_r{}^\mu{}_\nu\ ,}
\eqn\ksix
$$
where $\{J_r; r=1,2,3\}$ and 
$\{I_r; r=1,2,3\}$ are the quaternionic
structures on $\bR^4$ 
associated with the anti-self-dual and self-dual
two-forms, respectively. We remark that 
$$
[ J_r, I_s]=0\ .
\eqn\com
$$
We can arranged without loss of 
generality that  for this
intersection,
$\cMe$ is a hyper-K\"ahler manifold 
with torsion (HKT) with respect to 
pair $(\nabla^{(+)}, {\bf J}_r)$. This implies that
 the connection $\nabla^{(+)}$
has holonomy $Sp(2)$. The
metric on $\cMe$ can be written as
$$
ds^2_{(8)}\equiv \gamma_{ab} dx^a dx^b=
\big( U_{ij} \delta_{\mu\nu}+ V^r{}_{ij}
(I_r)_{\mu\nu}\big) dx^{i\mu} dx^{j\nu}\ ,
\eqn\hktmetr
$$
where the matrices $U, V^r$ have been given in [\pandrew] and 
$i,j=1,2$; $\mu, \nu=1,\dots,4$.
Due to \com, the above metric
is hermitian with respect to ${\bf J}_r$ 
as required. These geometries
are constructed by pulling-back with maps 
$$
\eqalign{
\tau:&\,\bH^2\rightarrow \bH
\cr
&\, q^i\rightarrow \tau(q^i)=p_i q^i-a}
\eqn\knine
$$
the four-dimensional HKT geometry 
associated with the NS-5-brane to $\bR^8$
and then summing up over the different choices of $\tau$,
where $q^i=x^{i1}+i x^{i2}+j x^{i3}+k x^{i4}$ and the 
quaternions $\{(p_i, a)\}$ are the
parameters of the maps.  The details of
the construction of the metric $ds^2_{(8)}$ 
and of the 3-form $H_{(8)}$ of 
this geometry will not
be given here. This has been done  
in [\pandrew] where it was
also found that the dilaton is
$$
e^{2\phi_{(8)}}= \gamma^{1\over4}\ .
\eqn\dilat
$$
The position of the NS-5-branes 
involved in the superposition
are given by the kernels of the 
above  maps $\tau$.
For later use we remark that the 
metric \hktmetr\ above has the property
$$
\gamma^{bc}\partial_a \gamma_{bc}=4 
\gamma^{bc}\partial_b\gamma_{ac}\ .
\eqn\keyprop
$$

To add a wave and a string to 
the above brane intersection,
we have to solve the equations for 
$g_1$ and $g_2$ in \feqgf.
Using \keyprop\ and the expression 
for the dilaton \dilat, we find that these
equations reduce to
$$
\eqalign{
\gamma^{i\mu, j\nu} \partial_{i\mu} \partial_{i\nu} g_1&=0
\cr
\gamma^{i\mu, j\nu} \partial_{i\mu} \partial_{i\nu} g_2&=0\ .}
\eqn\intstwaveh
$$
A class of solutions for these equations is
$$
g_1=1+\sum_{\{(p,a)\}}{\mu\big((p,a)\big)\over |p_i q^i-a|^2}
\eqn\intsolstwave
$$
and similarly for $g_2$. In order for a
wave and a string to lie on one
NS-5-branes, the parameters $\{(p,a)\}$ of 
$g_1$ and $g_2$ should be chosen to be the
same as those of the  maps $\tau$ used to 
construct the metric \hktmetr\ above.
However, for the most general solution of this type the
parameters $\{(p,a)\}$ of $g_1$, $g_2$ and the
metric can be different.

It remains to add rotation to the 
configuration by solving \feqgf\  for $F$.  Since $\cMe$ is a 
complex manifold with respect to three different 
complex structures $\{{\bf J}_r\}$,
 we shall seek solutions
for which $A$ is a hermitian-Einstein 
connection with respect to a
complex structure ${\bf K}$. An obvious 
choice for ${\bf K}$ is
as a linear combination of $\{{\bf J}_r\}$.  However, unlike
the situation in the flat case, here 
the hermitian-Einstein condition on $F$ 
with respect to such a choice of 
${\bf K}$
does {\sl not} imply the field 
equations \feqgf.
The difficulty lies in the presence of torsion.
To proceed,
we shall consider special
cases of the geometry \hktmetr\ 
above for which the manifold
$\cMe$ admits at least another 
K\"ahler structure 
with torsion (KT) with respect to
a pair $(\nabla^{(-)},{\bf  I})$. We can 
then show by direct computation
using the Jacobi Identities, the 
equation \keyprop\ and the 
 relation between the metric
and torsion in a KT geometry
that if
$$
F_{ab}=\Lambda (\omega_{{\bf  I}})_{ab}\ ,
\eqn\keleven
$$
then $F$ satisfies the field equations \feqgf, 
where $\Lambda$ is a constant.

There are two special cases that we 
shall consider the following:
$(i)\, $ We shall require that the 
geometry $\cMe$ admits another
KT structure with respect to the 
pair\foot{We can obviously 
instead of ${\bf  I}_3$ choose any of the
other two complex structures ${\bf  I}_1$ and 
${\bf  I}_2$ or even
a linear combination of all three 
complex structures. But all these
cases are symmetric, so by choosing ${\bf  I}_3$ 
there is no loss of
generality.} $(\nabla^{(-)},{\bf  I}_3)$.
If this is the case, then the holonomy 
of $\nabla^{(-)}$ is $SU(4)$.
To construct the metric and torsion of 
this geometry, we simply require
that the parameters $\{p_i; i=1,2\}$ of the 
linear maps $\tau$ are complex
 numbers instead of
quaternions. The metric \hktmetr\ becomes
$$
ds^2_{(8)}\equiv \gamma_{ab} dx^a dx^b=
\big( U_{ij} \delta_{\mu\nu}+ V_{ij}
(I_3)_{\mu\nu}\big) dx^{i\mu} dx^{j\nu}\ ,
\eqn\hktmetrb
$$
where  $V=V_3$ and $V_1=V_2=0$.
Choosing complex coordinates with respect
 to ${\bf I}_3$, the metric \hktmetrb\
is rewritten as
$$
ds^2_{(8)}\equiv \gamma_{ab} dx^a dx^b= 
Z_{ij}\delta_{\alpha \bar\beta} dz^{i\alpha}
dx^{j\bar\beta}= \big( U_{ij} + i V_{ij}\big)
\delta_{\alpha \bar\beta} dz^{i\alpha} dx^{j\bar\beta}\ .
\eqn\hktmetrc
$$
The hermitian-Einstein condition for 
${\bf K}={\bf I}_3$ implies that
$$
\eqalign{
F_{i\alpha, j\beta}&=0
\cr
Z^{ij} \delta^{\alpha \bar\beta}
 F_{i\alpha, j\bar\beta}&=0\ .}
\eqn\hemeistaa
$$
We again solve the first equation by setting
$$
\eqalign{
A_{i\alpha}&=\partial_{i\alpha} U
\cr
A_{i\bar\alpha}&=\partial_{i\bar\alpha}\bar U}
\eqn\kmone
$$
Substituting this in the second 
equation in \hemeistaa, we find that
$$
Z^{ij} \delta^{\alpha \bar\beta}
\partial_{i\alpha}\partial_{j\bar\beta}
\big(U-\bar U\big)=0
\eqn\kmtwo
$$
or
$$
\gamma^{i\mu, j\nu} \partial_{i\mu} 
\partial_{j\nu} \big(U-\bar U\big)=0
\eqn\intuone
$$
in real coordinates. This is precisely 
the equation that $g_1$ and $g_2$ for
the string and the wave satisfy, so it 
can be solved in a similar way.
To summarize, the solution is
$$
\eqalign{
ds^2&=2 g_1^{-1} dv(-du +A+g_2 dv)+ 
\big( U_{ij} \delta_{\mu\nu}+ V_{ij}
(I_3)_{\mu\nu}\big) dx^{i\mu} dx^{j\nu}
\cr
H&=-{1\over2} dv\wedge d(g_1^{-1} A)-
{1\over2} du\wedge dv\wedge dg_1^{-1}+ H_{(8)}
\cr
e^{2\phi}&= g_1^{-1} [\det(U_{ij} \delta_{\mu\nu}+ V_{ij}
(I_3)_{\mu\nu})]^{1\over4}\ ,} 
\eqn\ansa
$$
where
$$
\eqalign{
g_1&=1+\sum_{\{(p,a)\}}{\mu_1\big((p,a)\big)\over |p_i q^i-a|^2}
\cr
g_2&=\sum_{\{(p,a)\}}{\mu_2\big((p,a)\big)\over |p_i q^i-a|^2}
\cr
A_{i\mu}&= (I_3)^\nu{}_\mu \partial_{i\nu}
\big[\sum_{\{(p,a)\}}{\mu\big((p,a)\big)\over |p_i
q^i-a|^2}\big]\ ,}
\eqn\ihama 
$$
and $\{p_i\}$ are complex numbers. We remark that  
for the most general 
solution of this type the parameters
${\{(p,a)\}}$ of $g_1$, $g_2$ and of the metric can be 
different. The physical
interpretation of the solution above is that for every pair of 
parameters $(p,a)$ it corresponds 
a  wave with charge $\mu_2\big((p,a)\big)$ on string with
 charge $\mu_1\big((p,a)\big)$  both
  on a 5-brane with charge  $\mu_5\big((p,a)\big)$ located at
$$
p_iq^i-a=0\ .
\eqn\kmthree
$$ 
In addition each 5-brane has angular momentum
$$
{\cJ}= \mu\big((p,a)\big) I_3\ .
\eqn\kfour
$$
Notice that for every pair of 5-branes associated with the maps
 $(\tau, \tau')$ intersect on a string
located at
$\tau^{-1}(0)\cap \tau'^{-1}(0)$.  So the 
solutions has the interpretation of  waves
on  strings on rotating  
$Sp(2)$-intersecting NS-5-branes.

The fraction of spacetime
supersymmetry preserved by  solutions \ihama\
is summarized in the following table:

\vskip 0.3cm
$$
\vbox{\settabs 8\columns 
\+ {\rm Rotation}& $\surd$ & $-$&$-$ 
& $\surd$ & $\surd$ & $-$ & $\surd$
 \cr
\+ {\rm String}& $-$ & $\surd$&$-$ 
& $\surd$ & $-$ & $\surd$ & $\surd$
 \cr
\+ {\rm wave}& $-$ & $-$&$\surd$ & $-$ 
& $\surd$ & $\surd$ & $\surd$
 \cr
\+{\rm Susy} &${5\over 32}$& ${5\over32}$
&${3\over32}$& ${5\over32}$ & ${3\over32}$&
$~{3\over32}$&$~{3\over32}$\cr
\+{\rm Susy} &${1\over 16}$& $0$&${1\over16}$& $0$ & ${1\over16}$&
$~0$&$~0$\cr}
 $$
\vskip 0.2cm
The two rows with the fractions of 
supersymmetry correspond to the two possibilities
of letting the solutions of 
$\Gamma^{\underline v}\epsilon_+=0$ to be either in
$({\bf 1}_s,{\bf 8}_s)$ or in 
$({\bf 1}_c,{\bf 8}_c)$  subspaces of ${\bf 16}_s$,
respectively. For computing the 
above fractions of supersymmetry, we have  used
the orientation on  $\bE^8$ that it 
is induced from the complex structures ${\bf J}_r$
to write the chirality operator in 
eight-dimensions. Moreover observe that the
complex structures ${\bf I}_r$ give the same orientation on
$\bE^8$ as that of the complex structures 
${\bf J}_r$.  This is unlike the situation
in $\bE^4$ where the self-dual 
${ I}_r$ and anti-self-dual ${ J}_r$ complex structures
induce opposite orientations.

$(ii)\, $ We shall require that the 
geometry $\cMe$ admits another
HKT structure with respect to 
$(\nabla^{(-)},{\bf  I}_r)$. In this
case the holonomy of both connections 
$\nabla^{(-)}$ and $\nabla^{(-)}$ is $Sp(2)$.
To construct the metric and torsion of 
this geometry, we simply require
that the parameters $\{p_i; i=1,2\}$ of 
the linear maps $\tau$ are real
 numbers instead of
quaternions [\gtpap, \pandrew]. The metric \hktmetr\ becomes
$$
ds^2_{(8)}\equiv \gamma_{ab} dx^a dx^b= U_{ij} \delta_{\mu\nu}
 dx^{i\mu} dx^{j\nu}\ ,
\eqn\hktmetrb
$$
where  $V_1=V_2=V_3=0$.
Now we take $F$ to be in $Sp(2)$, i.e.
$$
F_{i\mu, j\nu} (I_r)^\mu{}_\rho (I_r)^\nu{}_\sigma
= F_{i\rho, j\sigma}
\eqn\sptwoo
$$
(no summation over r).
Since this is a special case of the 
Hermitian-Einstein condition, any connection
that satisfies \sptwoo\ also solves 
the field equations \feqgf.
The equation \sptwoo\ can be solved 
as in the flat case. So to summarize,
the solution in this case is
$$
\eqalign{
ds^2&=2 g_1^{-1} dv(-du +A+g_2 dv)+  
U_{ij} \delta_{\mu\nu} dx^{i\mu} dx^{j\nu}
\cr
H&=-{1\over2} dv\wedge d(g_1^{-1} A)-
{1\over2} du\wedge dv\wedge dg_1^{-1}+ H_{(8)}
\cr
e^{2\phi}&= g_1^{-1} \det(U_{ij})\ ,} 
\eqn\ansaaaa
$$
where
$$
\eqalign{
g_1&=1+\sum_{\{(p,a)\}}{\mu_1\big((p,a)\big)\over |p_i q^i-a|^2}
\cr
g_2&=\sum_{\{(p,a)\}}{\mu_2\big((p,a)\big)\over |p_i q^i-a|^2}
\cr
A_{i\mu}&= \sum_{\{(p,a)\}}\sum_{r=1}^3 (I_r)^\nu{}_\mu
\partial_{i\nu}{\mu^r\big((p,a)\big)\over |p_i q^i-a|^2}\ ,}
\eqn\gga
$$
and $\{p_i; 1,2\}$ are real numbers. The 
physical interpretation of this
solution is similar to that we have given 
for the solution \ansa\ of the previous
case. The angular momentum of each 5-brane 
involved in the intersection is
$$
\cJ=\sum_{r=1}^3 \mu^r\big((p,a)\big) I_r\ .
\eqn\kmfive
$$
The supersymmetry preserved by the solutions 
\ansaaaa\ is summarized in the following
table:
\vskip 0.3cm
$$
\vbox{\settabs 8\columns 
\+ {\rm Rotation}& $\surd$ & $-$&$-$ 
& $\surd$ & $\surd$ & $-$ & $\surd$
 \cr
\+ {\rm String}& $-$ & $\surd$&$-$ 
& $\surd$ & $-$ & $\surd$ & $\surd$
 \cr
\+ {\rm wave}& $-$ & $-$&$\surd$ 
& $-$ & $\surd$ & $\surd$ & $\surd$
 \cr
\+{\rm Susy} &${3\over 16}$
& ${3\over16}$&${3\over32}$& ${3\over16}$ & ${3\over32}$&
$~{3\over32}$&$~{3\over32}$\cr
\+{\rm Susy} &${3\over 32}$& $0$
&${3\over32}$& $0$ & ${3\over32}$&
$~0$&$~0$\cr}
 $$
\vskip 0.2cm
The two rows with the fractions of 
supersymmetry correspond to the two possibilities
of letting the solutions of 
$\Gamma^{\underline v}\epsilon_+=0$ to be either in
$({\bf 1}_s,{\bf 8}_s)$ or in 
$({\bf 1}_c,{\bf 8}_c)$  subspaces of ${\bf 16}_s$,
respectively. We remark that the same fractions of supersymmetry
are preserved by the associated toric hyper-K\"ahler
solutions of section three.

A special case of this class of solutions 
is to take the ratios $p_1/p_2$
of the parameters  $\{p_i; 1,2\}$ of 
all maps $\tau$ used to construct the
background 5-brane intersecting 
geometry to be the same. The resulting solution
has then the interpretation of 
parallel 5-branes. In this case, the holonomy
of the connections $\nabla^{(+)}$ 
and $\nabla^{(-)}$ is $sp(1)$. Moreover we can add
 rotations, waves and  strings as 
above. We may also take the ratios $p_1/p_2$
of the parameters $\{p_i; 1,2\}$ 
that appear in the expressions for $g_1$ and $g_2$ in \gga\
to be the same as those of the 
parameters of the  maps $\tau$ that are used in
the construction of the metric. 
The resulting solution will have the interpretation of
parallel 5-branes with  waves 
and  strings. The
supersymmetry preserved by this 
solution is summarized in the following table:

\vskip 0.3cm
$$
\vbox{\settabs 8\columns 
\+ {\rm Rotation}& $\surd$ & $-$&$-$ 
& $\surd$ & $\surd$ & $-$ & $\surd$
 \cr
\+ {\rm String}& $-$ & $\surd$&$-$ 
& $\surd$ & $-$ & $\surd$ & $\surd$
 \cr
\+ {\rm wave}& $-$ & $-$&$\surd$ 
& $-$ & $\surd$ & $\surd$ & $\surd$
 \cr
\+{\rm Susy} &${3\over 8}$& ${1\over4}$
&${1\over4}$& ${1\over4}$ & ${1\over4}$&
$~{1\over8}$&$~{1\over8}$\cr}
 $$
\vskip 0.2cm
The same fractions of supersymmetry are 
preserved for both choices of the projection
$\Gamma^{\underline v}\epsilon_+$.

There is also a third 
possibility for which  by suitably restricting
the parameters $\{p_i\}$ of the maps $\tau$  the
connection
$\nabla^{(-)}$ has holonomy ${\rm Spin}(7)$. 
This suggests that we may be able
 find rotations
for which $F$ is in ${\rm Spin}(7)$. We shall 
not further explore this here.

We finally remark that the solutions \ihama\ and \ansaaaa\  are not
 the most general ones
of the Killing and field equations with the above interpretation.
They should be rather thought as describing the asymptotic behaviour
 of more general solutions near 
the transverse spatial infinity.  It is expected
that these  more general solutions exist and this will 
be exploited to
investigate the geometry near the NS-5-brane intersections
in section eight.

\chapter{Rotations and Ricci Flat Cones}

So far we have chosen eight-dimensional
manifolds ${\cal M}_{(8)}$  that
are asymptotically  flat. However, our ansatz allows
many other choices of ${\cal M}_{(8)}$. 
One of them is that of a Ricci-flat cone
$C({\cal N}_{(7)})$ over a 
seven-dimensional manifold ${\cal N}_{(7)}$. 
To preserve some supersymmetry, the 
manifolds ${\cal N}_{(7)}$ are 
chosen such that their associated cones
  have holonomy $SU(n)$,
 $n=2,3,4$ (Calabi-Yau), $Sp(2)$ 
(hyper-K\"ahler),
$G_2$ and ${\rm Spin}(7)$. Moreover 
the dilaton $\phi_{(8)}$ is constant
and the three-form field strength 
$H_{(8)}$ vanishes. The metric 
on ${\cal M}_{(8)}$ is 
$$
ds^2_{(8)}=dr^2+r^2 ds^2_{(7)}\ ,
\eqn\kmsix
$$
where 
$$
ds^2_{(7)}=h_{ij} dy^i dy^j
\eqn\kmseven
$$ 
is the metric on ${\cal N}_{(7)}$ and 
$r$ is a radial coordinate. 
In many cases, ${\cal M}_{(8)}$ are 
singular at $r=0$. Such eight-dimensional
geometries have recently appear  in the 
investigation of near horizon geometries
of the M-2-brane and of the M-5-brane [\hull]. 

A  class of solutions with a string and a wave 
can be found by assuming that $g_1$ and
$g_2$ are functions of the radial coordinate $r$. In this
 case, we have that
$$
\eqalign{
g_1&=1+{m_1\over r^6}
\cr
g_2&={m_2\over r^6}\ .}
\eqn\chamm
$$
Obviously more general solutions can be found by 
taking $g_1$ and $g_2$ to be
general harmonic functions on $C({\cal N}_{(7)})$ and 
therefore to depend on the coordinates
of ${\cal N}_{(7)}$ as well.
The equation of motion for $F$ can be rewritten as
$$
\eqalign{
\partial_j\big( {\sqrt h} h^{jk} F_{kr}\big)&=0
\cr
\partial_j\big( r^3 {\sqrt h} h^{jk} h^{i\ell} 
F_{k\ell}\big)+\partial_r\big( r^5 {\sqrt h}
h^{i\ell} F_{r\ell}\big)&=0\ .}
\eqn\hone
$$
Now if $F_{ri}=0$ and $F_{ij}=F_{ij}(y)$, then
 $F$ is a harmonic 
two-form on ${\cal N}_{(7)}$.
Such solutions are then associated 
with principal $U(1)$ bundles $\cL$ over
${\cal N}_{(7)}$. For trivial bundles, 
the rotation can be eliminated
with a coordinate transformation\foot{We assume that the
connection is also trivial.}. So 
the interesting cases are those
associated with non-trivial bundles. 
This requires that the second betti number of
${\cal N}_{(7)}$ to be non-vanishing. 
Many such examples have been constructed.
These include the tri-Sasakian manifolds 
of [\trisasak] which give rise to hyper-K\"ahler
cones, and their associated by squashing 
weak $G_2$ holonomy seven-dimensional
manifolds which give rise to cones 
with ${\rm Spin}(7)$ holonomy. Other examples 
include some of the Sasaki-Einstein 
spaces of [\saseins] which are associated
with Calabi-Yau cones. If $A$ vanishes our solutions 
specialize those considered in [\ital, \hull].

It is clear that as $r\rightarrow \infty$, 
our solutions
 are  asymptotically $\tilde\cL\times \bR$, 
where $\tilde\cL$ is a 
 principal $U(1)$ bundle
over $C({\cal N}_{(7)})$ induced 
from the principal
 $U(1)$ bundle on  ${\cal N}_{(7)}$. Since these spacetimes
are not asymptotically flat,  $A$ is not
straightforwardly related to angular 
momentum at infinity. The fractions supersymmetry
preserved by our solutions are 
summarized in the following table: 

\vskip 0.3cm
$$
\vbox{\settabs 7\columns 
\+ {\rm Holonomy}& $SU(2)$ & $SU(3)$&$SU(4)$ 
& $Sp(2)$ & $G_2$ & ${\rm Spin}(7)$ 
 \cr
\+ {\rm Susy}& $~{1\over8}$ & $~{1\over16}$&$~{1\over16}$ 
& ${3\over32}$ & ${1\over16}$ & $~{1\over32}$ 
 \cr
\+ {\rm Susy}& $~{1\over8}$ & $~{1\over16}$&$~~0$ 
& $~0$ & $~0$ & $~~0$ 
 \cr}
 $$
\vskip 0.2cm
For the calculation of the above fractions,
we have assumed that $F$ is generic 
and so the Maxwell Killing
spinor equation does not admit any 
non trivial solutions. In this case
the above fractions do not alter with 
the addition of strings and waves
in the configuration. The two supersymmetry 
rows correspond to the choice
of taking the solutions of 
$\Gamma^{\underline v}\epsilon_+=0$ to be either
in $({\bf 1}_s, {\bf 8}_s)$ or in 
$({\bf 1}_c, {\bf 8}_c)$, respectively.

\chapter{M-theory}

The  solutions that we have described 
in the previous  sections
as solutions of IIA string theory  
can be easily  lifted
to M-theory. For this  let $z$ be the 
eleventh coordinate. The relevant Kaluza-Klein 
ansatz\foot{We have not included in this ansatz 
the Kaluza-Klein vector and the IIA four-form 
field strength because they vanish for our
ten-dimensional solutions.} for the reduction 
from eleven dimensions to ten is
$$
\eqalign{
ds^2_{(11)}&= e^{{4\over3}\phi} 
dz^2+ e^{-{2\over3}\phi} ds^2_{(10)}
\cr
G_4&=H\wedge dz\ ,}
\eqn\hhone
$$
where $\phi$ is the ten-dimensional dilaton, 
the ten-dimensional metric $ds^2_{(10)}$ is in the
string frame, $G_4$ is the 4-form field
 strength of eleven-dimensional
 supergravity and $H$
is the ten-dimensional 
NS$\otimes$NS three-form field strength. The lifting
to M-theory of solutions 
given by the ansatz \ans\ is
$$
\eqalign{
ds_{(11)}^2&= g_1^{-{2\over3}} 
\big[ e^{{4\over3}\phi_{(8)}} dz^2+ 2 
e^{-{2\over3}\phi_{(8)}} 
dv(-du +A+g_2 dv)\big]+ g_1^{{1\over3}}
e^{-{2\over3}\phi_{(8)}} ds^2_{(8)}
\cr
G_4&=\big [-{1\over2} dv\wedge 
d(g_1^{-1} A)-{1\over2} du\wedge dv\wedge dg_1^{-1}+
H_{(8)}\big ]\wedge dz}
\eqn\mans
$$
for $\lambda=-{1\over2}$.
Since this solution of M-theory is 
constructed from lifting a solution of IIA
string theory is not localized in the 
$z$ direction. It turns out that we can modify
our ansatz in eleven dimensions to allow 
localization of  $g_2$ associated with the
wave as it has been done in the special 
case investigated in [\jerometwo]. However in what follows
this property of $g_2$ will not be used 
and so we shall not elaborate further on
this point.

It is well known that IIA strings lift to 
M-theory as  M-2-branes,  pp-waves lift
again as  pp-waves and NS-5-branes 
lift as M-5-branes. Moreover it turns out
that the rotation associated with the abelian 
field-strength $A$ also lifts as
a rotation. The worldvolume coordinates of the 
M-2-brane are $(u,v,z)$.  Following this
IIA/M-theory duality correspondence 
interpretation of the lifted solutions of the IIA
solutions that we have found in the previous 
sections is straightforward. We remark that it is
not necessary to localize $A$ in the 
$z$ coordinate. This is because
 the number of
transverse directions of the M-2-brane is  
the same as number of transverse 
directions of the IIA string.
For example the wave on a rotating string 
 solution of section (3.5) is lifted
to M-theory as
$$
\eqalign{
ds_{(11)}^2&= g_1^{-{2\over3}} \big[ dz^2+ 2 
 dv(-du +A+g_2 dv)\big]+ g_1^{{1\over3}}
 ds^2(\bE^8)
\cr
G_4&=-{1\over2}\big [ dv\wedge d(g_1^{-1} A)+ 
du\wedge dv\wedge dg_1^{-1}\big]\wedge dz\ ,}
\eqn\mrsw
$$
where $g_1$ and $g_2$ are harmonic functions on 
$\bE^8$ and the field strength $F$ of $A$ is
in $su(4)$ or in $spin(7)$ or in $so(8)$. The 
explicit expressions of $A$ in all three
cases have
been given in sections (3.1), (3.3) and (3.4), 
respectively. The angular momentum is the
same as that we have computed  in the 
context of IIA strings. 

As another example, we  lift  the solutions 
for which $\cMe$ is a Ricci-flat cone of
section six. The resulting M-theory solution is
$$
\eqalign{
ds_{(11)}^2&= g_1^{-{2\over3}} \big[ dz^2+ 2 
 dv(-du +A+g_2 dv)\big]+ g_1^{{1\over3}}
 \big(dr^2+r^2 ds^2_{(7)}\big)
\cr
G_4&=-{1\over2}\big [ dv\wedge d(g_1^{-1} A)+ 
du\wedge dv\wedge dg_1^{-1}\big]\wedge dz\ ,}
\eqn\mcrsw
$$
where $g_1$ and $g_2$ are given in \chamm\ 
and $A$ is a connection of a principal $U(1)$
bundle  on
${\cal N}_{(7)}$ for which $F$ is harmonic.

Finally, the rotating rotated intersecting 
branes of section five are lifted as follows:
$$
\eqalign{
ds_{(11)}^2&= g_1^{-{2\over3}} 
\big[ e^{{4\over3}\phi_{(8)}} dz^2+ 2 
e^{-{2\over3}\phi_{(8)}} dv(-du +A+g_2 dv)\big]+
\cr & g_1^{{1\over3}}
e^{-{2\over3}\phi_{(8)}} 
\big( U_{ij}\delta_{\mu\nu}+V^1_{ij} (I_3)_{\mu\nu}\big) dx^{i\mu}
dx^{j\nu}
\cr
G_4&=\big [-{1\over2} dv\wedge d(g_1^{-1} A)
-{1\over2} du\wedge dv\wedge dg_1^{-1}+
 H_{(8)}\big ]\wedge dz\ ,}
\eqn\mrrib
$$
where $e^{2\phi_{(8)}}$ is 
given \dilat, and   $g_1, g_2$ and $A$ 
are given in \ihama\ in case
(i) or in
\gga\ for case (ii). For $A=0$, the 
lifting has been described in [\pandrew].

\chapter{Near Horizon Geometries}

\section{$Sp(2)$-Intersecting NS-5-branes}

We shall first consider the near horizon 
geometry of two NS-5-branes 
 intersecting  at angles on a 
string associated with 
the maps $\tau_1$ and $\tau_2$,
respectively.
Moreover let us assume that
$\tau_1^{-1}(0)\cap
\tau_2^{-1}(0)=\{x\}$ is a point in $\bH^2$; in such case the
two NS-5-branes intersect on string. Next, 
we change coordinates in $\bH^2$ as
follows:
$$
\eqalign{
x&=\tau_1(q^i)
\cr
y&=\tau_2(q^i)\ .}
\eqn\htwo
$$
This change of coordinates  
is an invertible transformation.
The metric and three-form field strength in 
these new coordinates can be re-expressed as
$$
\eqalign{
ds^2&=ds^2(\bE^{(1,1)})+ds^2_\infty
+{R^2_1\over |x|^2} (|dx|^2+ |x|^2 ds^2(S^3))
\cr &
+
{R^2_2\over |y|^2} (|dy|^2+ |y|^2 ds^2(S^3))
\cr
H&= R^2_1 {\rm Vol}(S^3)+R^2_2 {\rm Vol} (S^3)\ ,}
\eqn\twobranehor
$$
where ${\rm Vol}(S^3)$ is the volume form of $S^3$, $ds^2_\infty$ is 
the constant asymptotic
metric at the transverse spatial infinity,  and
$R^2_1=\mu(\tau_1)$ and $R^2_2=\mu(\tau_2)$ are constants. Now in the 
limit that both 
$|x|^2<< R^2_1$ and  
$|y|^2<< R^2_2$, the constant
asymptotic part
of the metric can be neglected and the near
 horizon geometry is $\bE^{(1,3)}\times S^3\times
S^3$. This near horizon geometry  is exactly the
 same as that one finds for orthogonally 
intersecting NS-5-branes for which 
 the asymptotic constant
metric is
$ds^2_\infty=|dx|^2+|dy|^2$.

Next consider the case where more than 
two branes are involved in the intersection.
It suffices to consider the  case  of three intersecting
 branes associated with the 
maps $\tau_1$, $\tau_2$ and $\tau_3$,
respectively. This is because the
arguments that we shall present to 
determine the near horizon 
geometry in this case can be easily
extended to N branes.  We 
shall take that the sets  $\tau_1^{-1}(0)$,
$\tau_2^{-1}(0)$ and 
$\tau_3^{-1}(0)$ pairwise to intersect on a point and 
$\tau_1^{-1}(0)\cap \tau_2^{-1}(0)\cap\tau_3^{-1}(0)=\emptyset$, i.e. 
in differential topology
terminology the three branes are 
in general position\foot{This assumption
appears to be necessary because  
otherwise the metric is singular.}. 
In particular, the latter condition implies
that not all three branes 
intersect on the same string.  Now as
we approach the intersection 
region of the branes associated
 with say the maps $\tau_1$ and $\tau_2$,
the metric is dominated by the $x$ and $y$ coordinates 
adapted to these two branes. This is
because by assumption the intersections of all 
three branes are well separated.
So the near horizon geometry at the intersection of 
the above pair of branes is $\bE^{(1,3)}\times
S^3\times S^3$ as for the two 
brane intersection above, and similarly 
for the near horizon geometry at the intersection 
of the other two  pairs.

\section{Rotating $Sp(2)$-Intersecting NS-5- and M-5-branes}

To find a smooth geometry near the 
pairwise $Sp(2)$-intersection of rotating
  NS-5-branes superposed with strings
and waves, we have to seek for more general  solutions 
from those found in section (5). For these new solutions,
 the functions $g_1, g_2$ associated with the  string and the wave, 
 and  the $U(1)$ gauge field
$A$ are again solutions of
\intstwaveh\ and \intuone, respectively, but they
 should have different asymptotic
behaviour near the  
intersection regions  from those  of \ihama.
Unfortunately, we were not able to solve the 
equations \intstwaveh\ and \intuone\  exactly.
So to achieve our purpose we shall construct 
an approximate solution near the horizon
which has the desirable behaviour. This is related to the 
observation in [\arkadytwo]
that generalized harmonic function
equations simplify near horizons.

We shall first consider strings and waves in the background of 
two $Sp(2)$-intersecting
NS-5-branes on a string.
To proceed we shall first estimate the behaviour of 
the inverse of the metric $\gamma$ of
$\cMe$ associated with the solution
\twobranehor\ in the limit $|x|^2<< R^2_1$ and  $|y|^2<< R^2_2$. It 
is easy to see that
schematically
$$
\gamma^{-1}=\gamma_h^{-1}+O\big(\big({r\over R}\big)^4\big)
\eqn\hthree
$$
where  $\gamma_h^{-1}$ is the inverse of the metric
$$
ds^2_h={R^2_1\over |x|^2} (d\bar x dx+ |x|^2 ds^2(S^3))
+
{R^2_2\over |y|^2} (d\bar y dy+ |y|^2 ds^2(S^3))\ ,
\eqn\hfive
$$
i.e. the inverse of the metric near the 
intersection, and $(r/R)^2$ denotes  ratios
of the type
$(|x|/ R_1)^2, (|y|/ R_2)^2, (|x||y|/ R_2 R_1)$. 
Using this we can estimate the solutions of 
$g_1$, $g_2$ and $A$ of \intstwaveh\ and
\intuone\ in the same limit. In particular, a solution  for $g_1$ is
$$
g_1= \mu_1{R^2_1 R^2_2\over r_1^2 r_2^2}+O\big(\big({R\over r}\big)^2\big)
\eqn\appsol
$$
and similarly for $g_2$, where $r_1=|x|$ and $r_2=|y|$. A solution for $A$ is
$$
A=\mu{R^2_1 R^2_2\over r_1^2 r_2^2} \big(\sigma_3+\tilde \sigma_3\big)
+O\big(\big({R\over r}\big)^2\big)
\eqn\asypcone
$$
where the K\"ahler form ${\bfomega}_t$ of the 
complex structure ${\bf I}_t$ is
$$
{\bfomega}_t={1\over2} [d(r_1^2\sigma_t+r_2^2\tilde\sigma_t)]\ ,
\eqn\asta
$$
and
$$
\eqalign{
d\sigma_t&=\epsilon_{tsp}\sigma_s\wedge \sigma_p 
\cr
d\tilde\sigma_t&=\epsilon_{tsp}\tilde\sigma_s\wedge \tilde\sigma_p }
\eqn\valta
$$
are left invariant one-forms on $S^3\times S^3$.
It is clear that in both the above cases the first order term dominates 
in the limit $r_1^2<<R^2_1$ and  $r_2^2<< R^2_2$.
So  the metric \ansa\ becomes
$$
\eqalign{
ds^2\sim& -2\mu_1{ r_1^2\over R^2_1} {r_2^2\over R^2_2 } dvdu+
2\mu\mu_1  dv (\sigma_3+\tilde \sigma_3)
\cr &+
2\mu_2 \mu_1 dv^2+ {d^2r_1\over r_1^2}+
{d^2r_2\over r_2^2}+ds^2(S^3)+d\tilde s^2(S^3)\ .}
\eqn\nearhmetr
$$
After a redefinition of the various constants,  
the near horizon geometry of 
strings and waves on two rotating
$Sp(2)$-intersecting NS-5-branes on a string 
is the same as that of the associated
orthogonal intersection investigated in [\cowdall, \boonstra].
So we find after a coordinate 
transformation that the metric \nearhmetr\
is  $AdS_3\times S^3\times S^3\times \bE$.
In general the  geometry near the intersection 
of every pair of waves on strings on $N$
rotating
$Sp(2)$-intersecting NS-5- branes, for which the NS-5-branes
are in general position, is  
$AdS_3\times S^3\times S^3\times \bE$. This follows from the more
detail analysis of previous section. We 
remark that it appears that the approximate solutions
\appsol\ above have the appropriate decay as 
$|x|^2>> R^2_1$ and  $|y|^2>> R^2_2$ to match
the solutions given in \ihama\ at the spatial 
transverse infinity region. However a more 
detail analysis is needed to establish
this.

As we have seen this solution is lifted to 
M-theory to another solution with
the interpretation of membranes ending on  rotating
$Sp(2)$-intersecting M-5-branes  superposed with waves.
The investigation of the near horizon geometry 
in this case is very similar to
the one presented above. So we shall 
not repeat the analysis here.
In particular we find that near the 
intersection of every pair of M-5-branes
the geometry is $AdS_3\times S^3\times ^3\times E^2$.
In fact it is the same as the near horizon 
geometry of the associated  orthogonal
intersection investigated in [\jerometwo].

\chapter{Null Compactifications }

Compactifications of  M-theory 
(or strings) to $11-k$ dimensions
 involve supergravity solutions
of the form $\bE^{(1, 10-k)}\times {\cal N}_{(k)}$,
where ${\cal N}_{(k)}$ is a compact manifold. A large class of 
such solutions can be given by taking 
 ${\cal N}_{(k)}$ to be one of the 
manifolds with special holonomy. These
include manifolds with holonomy $SU(n)$,
 $n=2,3,4$ (Calabi-Yau), $Sp(2)$ (hyper-K\"ahler),
$G_2$ and ${\rm Spin}(7)$. Such manifolds ${\cal N}_{(k)}$ solve 
the supergravity field
equations and preserve some of the spacetime supersymmetry 
provided that we take the various
form  field strengths to vanish and the
dilaton to be constant.

Our ansatz \ans\ allows the construction of  
solutions of strings and M-theory
associated with new compactifications that
 have the topology of a $U(1)$ principal bundle $\cL$ over 
the manifolds associated with
above standard compactifications. However in this case,
 the compact direction along
the fibre direction of $\cL$ is
null and the string or membrane  form field strength 
gets an non-zero expectation value.
Null compactifications of gravity theories has been 
investigated in the past, see for example in [\julia]. Here we propose
an adaptation of these compactifications in the context of 
strings and M-theory. 
To find such solutions, we set $g_1=1$ and $g_2=0$ in 
our ansatz \ans. In addition,
let 
 $\cL$ be a principal $U(1)$ bundle over 
${\cal N}_{(k)}$ equipped with a
connection $A$ with curvature $F$. Then, the manifold 
$\cMe$ has topology 
$\bR^{7-k}\times \cL$. The field equations
require that $F$ is a harmonic two-form on
 ${\cal N}_{(k)}$. Now if $\cL$ is
topologically trivial, then $F=0$ and the 
solution is $\bR^{9-k}\times S^1\times {\cal N}_{(k)}$.
So for non-trivial solutions we should take 
$\cL$ to be topologically non-trivial. 
Supersymmetry imposes additional conditions on 
$F$ but these depend on the
properties of ${\cal N}_{(k)}$ and $\cL$ and they will 
be investigated below.

\section{Null Toric Compactifications}

The simplest case to consider  is ${\cal
N}_{(k)}=T^k$.  Let $\{x^a; a=1,\dots,k\}$ be the 
periodic angular coordinates on $T^k$
with periodicities $\{r^a; a=1,\dots, k\}$, respectively.
We write the metric of $T^8$ as
$$
ds^2=\gamma_{ab} dx^a dx^b\ ,
\eqn\hsix
$$
where $\{\gamma_{ab}\}$ is a constant matrix.
The field equations for $A$ imply that 
$F={1\over2}F_{ab} dx^a\wedge dx^b$ is constant. This
in turn implies that
$$
A_a= F_{ab} x^b+c_a\ ,
\eqn\hseven
$$
where $c$ is a constant. Given a basis 
$\{C_a; a=1,\dots, k\}$ of $H_1(T^k, \bZ)$
adapted to  this coordinate system, we find the 
dual basis $\{\lambda^a; a=1,\dots, k\}$ in
$H^1(T^k, \bZ)$, where
$$
\lambda^a={1\over r^a} dx^a\ ;
\eqn\hnine
$$ 
(no summation over the index  $a$). 
A basis in $H^2(T^k, \bZ)$ can 
then be found by taking
$$
\lambda^{ab}=\lambda^a\wedge \lambda^b
\eqn\hten
$$
for $a<b$. Rewriting $F$ in this basis, we get
$$
F=f_{ab}\lambda^{ab}\ .
\eqn\hmone
$$
For $F$ to be the curvature of a line bundle it is 
required that all the constants
$f_{ab}$ are integers. This gives a large class of 
supergravity solutions with rotation.
For generic choices of $F$ such solutions  
preserve $1/4$ of supersymmetry.
There are special choices of $F$ for
 which more supersymmetry is preserved
by the solution. For example, one can 
introduce a complex structure on $T^k$, $k$ even,
and look for connections that satisfy 
the Hermitian-Einstein condition.
There are many such connections which 
will lead to solutions that will
preserve more supersymmetry. We shall 
not further elaborate on this point here.

\section{Null Calabi-Yau Compactifications}

On manifolds with at least one 
complex structure the most natural
supersymmetry condition to impose 
on $F$ is to require that $A$ is
a Hermitian-Einstein condition. 
Let ${\cal N}_{(k)}$ be a Calabi-Yau
manifold with metric $\gamma$ 
and complex structure $I$. 
The Hermitian-Einstein condition becomes
$$
\eqalign{
F_{\alpha\beta}&=0
\cr
\gamma^{\alpha\bar\beta} F_{\alpha\bar\beta}&=0\ ,}
\eqn\hermeinb
$$
in complex coordinates with respect to 
$I$. Moreover, the above two conditions
imply that $F$ satisfies the Maxwell 
equations on ${\cal N}_{(k)}$.
It is well known that the first 
condition has solutions provided that
the $\cL$ is holomorphic. So it remains 
to find the conditions required for the second equation 
in \hermeinb\ to have solutions. For this we define the k-form
$$
\lambda=\omega_I^{{k\over2}-1}\wedge c_1(\cL)\ ,
\eqn\hmthree
$$
where $\omega_I$ is the K\"ahler 
form of $I$ and $c_1(\cL)$ is 
the first Chern class of 
$\cL$. It turns out that a necessary 
and sufficient condition for the existence
of a connection that satisfies the 
second equation in \hermeinb\ is  that
$$
\int_{{\cal N}_{(k)}}\lambda=0\ ;
\eqn\hermeinstcond
$$
(see for example [\green] and references within).
It would be of interest to investigate 
the small fluctuations
of string and M-theory around such
 compactifications. The fraction 
of supersymmetry preserved
by such Null compactifications is the 
same as that of the associated
Calabi-Yau ones.

These considerations can be extended to 
compactifications for which ${\cal N}_{(k)}$
is hyper-K\"ahler, $G_2$ or ${\rm Spin}(7)$ [\paptown]. 
It is natural to impose that $F$ is in
$sp({k\over4})$, $g_2$ or $spin(7)$, 
respectively. It is clear that to find
solutions to these equations,  $\cL$ must be 
restricted appropriately
in a way similar to the 
Hermitian-Einstein case in  \hermeinstcond.

\chapter{Concluding Remarks}

We have investigated the various ways of adding angular momentum
to branes preserving supersymmetry. The resulting configurations
have the interpretation of rotating branes. 
We have applied our methods to 
added rotation to $Sp(2)$-intersecting NS-5-branes on a string 
with  superposed strings
and pp-waves. In this way,  we constructed new solutions with 
the interpretation of
rotating rotated branes. Superpositions 
with strings, waves and membranes were
also considered where appropriate.
We then explored the related M-theory 
configurations. We also found  approximate
solutions of rotating rotated NS-5-branes in 
type II strings and of rotating rotated  M-branes in M-theory 
with near horizon geometries
$AdS_3\times S^3\times S^3\times \bE$ and
$AdS_3\times S^3\times S^3\times \bE^2$, respectively.  
Our results can also be adapted in the 
context of D-branes by using T- and S-duality.
Moreover, we have presented  compact solutions that
can be used for null compactifications of strings and M-theory.

An application of our solutions is in the 
context of five-dimensional
black holes. Since in five dimensions a 
three-form field strength is dual
of a two-form one. We can first reduce many 
of our solutions to five dimensions
and then dualize the three-form field strength. 
The resulting solutions
will be a supersymmetric rotating black hole 
solution with various $U(1)$ charges.
It would be of interest to investigate these 
five-dimensional black hole
solutions in the future. It is also well 
known that there is a correspondence between
supergravity solutions and brane worldvolume 
solitons. If this correspondence
holds in this case, then there must be 
worldvolume solutions that have
the interpretation rotating branes and 
preserve the same fraction of
supersymmetry as that of the 
associated supergravity solutions.

A large class of supergravity 
solutions has been constructed
by superposing or intersecting a 
small number of \lq elementary' ones
using some superposition or intersection rules.
These superpositions can involve a 
large number of \lq elementary'
solutions. The success of this method 
can be easily demonstrated
by the increasing complexity and variety of 
the superposed solutions. However further
progress
towards understanding the supergravity solutions
will depend on the number and interpretation 
of \lq elementary' solutions, as
well as the development of geometric methods  to treat manifolds that
are equipped with natural k-forms. Global charges and their
 algebras may be useful to find
the \lq elementary' solutions [\townsenda, \hullb]. However 
no all global charges  appear
in the supersymmetry algebra, like for example the angular momentum
for asymptotically flat spacetimes. Since most of the solutions that have
been found
preserve a proportion of spacetime 
supersymmetry, they admit connections,
some with torsion, that have special holonomy. In this respect they
are closely related to well known special holonomy
 manifolds like the hyper-K\"ahler
and Calabi-Yau ones. If this is the case, 
then this correspondence indicates that
 many of the
solutions that have been constructed 
represent special points in the moduli space
of all possible solutions with the same 
holonomy and similar interpretation. Some evidence 
that this may be the case has been provided in [\mukhi]. However 
since in the context
of branes
connections that appear are not the Levi-Civita ones many of
the key properties of the standard special holonomy
manifolds do not generalize.  Nevertheless, it is likely that new methods can
 be developed in the near future 
to investigate many of the properties of supergravity solutions.

\vskip 1cm
\noindent{\bf Acknowledgments:}    I would like 
to thank P.K. Townsend for suggesting 
the possibility of rotating rotated branes and 
G.W. Gibbons for helpful discussions
on manifolds with exceptional holonomy. I am
supported by a University Research Fellowship from the Royal Society.

\appendix
\centerline{T-DUALITY, D-BRANES AND ROTATIONS}

The type II  T-duality and type IIB  S-duality can be used  to construct
rotating D-branes   beginning from the 
rotating fundamental IIB string 
of section (3.5). For this we assume that 
the original solution is independent
from the compactifying coordinate $x$ that the 
T-duality is performed and that the $U(1)$
gauge potential $A$
associated with the rotation does not have components along $x$.  For
example performing one S-duality and two 
T-dualities on the fundamental IIB string, we find
the metric of a rotating D-3-brane  as follows:
$$
ds^2= g_1^{-{1\over2}} \big( 2 dv (-du+A)+ds^2(\bE^2)\big) + 
g_1^{{1\over2}} ds^2(\bE^6)\ ,
\eqn\dthreerot
$$
where $g_1$ is a harmonic function on 
$\bE^6$ and $A$ is a solution of the Maxwell
field equations in $\bE^6$. Such 
solutions for $A$ have been given in section (3).
In particular, one such  solution is that 
given  by requiring that $F$ is  in
$su(3)$. The D-3-brane then has 
angular momentum proportional to the complex
structure associated with the 
embedding of $su(3)$ in $so(6)$. A preliminary 
computation has revealed
that the  near
horizon geometry of this rotating D-3-brane 
solution may be singular. However further
investigation is required to establish this.

In addition T-dualizing twice and S-dualizing 
once the above rotating D-3-brane, we find a new
rotating 
NS-5-brane solution with
metric
$$
ds^2=  \big( 2 dv (-du+A)+ds^2(\bE^4)\big) + g_1 ds^2(\bE^4)\ ,
\eqn\dthreerot
$$
where $g_1$ is harmonic 
function in $\bE^4$ and $A$ satisfies the Maxwell
equations. This appears to contradict the field 
equations for $F$ found in section (2).
However this is not the case. The 
resolution of this puzzle is that there is a
more general ansatz than the one we 
have used to add rotation to  branes. 
This  will be investigated in the future.

\refout

\bye